\documentclass[preprint,review,12pt]{elsarticle}

\usepackage{graphicx}
\usepackage{subcaption}
\usepackage{array}
\usepackage{tabularx}
\usepackage{amsmath}
\usepackage{setspace}

\DeclareMathOperator*{\argmax}{argmax}

\newcommand{\PreserveBackslash}[1]{\let\temp=\\#1\let\\=\temp}
\newcolumntype{C}[1]{>{\PreserveBackslash\centering}p{#1}}

\journal{Signal Processing}

\begin{document}

\begin{frontmatter}



\title{\vspace*{-2.0cm}Maximum Likelihood Speed Estimation of Moving Objects in Video Signals}


\author{Veronica Mattioli\corref{cor1}}
\cortext[cor1]{Corresponding author}
\ead{veronica.mattioli@unipr.it}
\author{Davide Alinovi\fnref{label2}}
\fntext[label2]{Deceased on 16 September 2020}
\author{Riccardo Raheli}
\ead{riccardo.raheli@unipr.it}

\address{Department of Engineering and Architecture, University of Parma, Parco Area delle Scienze 181/A, IT-43124 Parma, Italy \vspace*{-1cm}}


\begin{singlespace}
\begin{abstract}
Video processing solutions for motion analysis are key tasks in many computer vision applications, ranging from human activity recognition to object detection. In particular, speed estimation algorithms may be relevant in contexts such as street monitoring and environment surveillance. In most realistic scenarios, the projection of a framed object of interest onto the image plane is likely to be affected by dynamic changes mainly related to perspectival transformations or periodic behaviours. Therefore, advanced speed estimation techniques need to rely on robust algorithms for object detection that are able to deal with potential geometrical modifications. The proposed method is composed of a sequence of pre-processing operations, that aim to reduce or neglect perspetival effects affecting the objects of interest, followed by the estimation phase based on the Maximum Likelihood (ML) principle, where the speed of the foreground objects is estimated. The ML estimation method represents, indeed, a consolidated statistical tool that may be exploited to obtain reliable results. The performance of the proposed algorithm is evaluated on a set of real video recordings and compared with a block-matching motion estimation algorithm. The obtained results indicate that the proposed method shows good and robust performance.
\end{abstract}

\begin{keyword}
	maximum likelihood \sep speed estimation \sep video signals \sep perspectival transformations
	
	
	
\end{keyword}

\end{singlespace}
%
%

\end{frontmatter}


\section{Introduction}\label{intro}
Speed estimation systems play nowadays a fundamental role in contexts of traffic control and road monitoring applications. Thanks to the increasing deployment of cameras for surveillance purposes, a significant amount of street-related information is available and may be exploited to build non-intrusive video-based solutions for object speed estimation.

In many realistic scenarios, the motion of foreground objects is superposed to other dynamic modifications that mainly arise from periodic behaviours (e.g., typical of some human movements such as walking and running) or directly result from the process of image acquisition \cite{Solomon}, \cite{Richard}. A video frame can be indeed defined as a digital image generated by the projection of a three-dimensional (3D) real-world scene onto a two-dimensional (2D) camera plane. For this reason, perspectival effects are likely to affect the image of the framed objects of interest. Hence the analysis of their motion, i.e., speed estimation and periodic feature extraction, may be challenging in some scenarios.

The goal of this paper is the estimation of the speed of foreground objects in a video sequence, accounting for the perspectival effects that may affect their 2D projection throughout the video duration. To this purpose, we may apply inverse projective transformations \cite{Solomon}, \cite{ViewGeom} to each frame of the video sequence to obtain a new processed sequence where the shape and the size of the framed objects are affected by minor modifications and can be considered constant throughout the video duration. We exploit a sound estimation theory tool, i.e. the Maximum Likelihood (ML) principle \cite{Kay} to obtain an estimator of the speed of the objects of interest.

An extensive review of video-based solutions for speed estimation can be found in \cite{FernndezLlorca2021VisionbasedVS}, where various methods are classified according to specific criteria. In particular, classical approaches can be mainly categorized as based on a) camera setting and calibration and b) vehicle detection and tracking. The works described in \cite{SOCHOR201787} and \cite{Giannakeris_2018_CVPR_Workshops} are examples in the first category, where vanishing points are detected on the considered 2D scene and the camera parameters are computed to enable the measurement of distances on the road plane, hence the speed of the vehicle of interest. On the other hand, solutions based on category b) vehicle detection and tracking, mainly rely on background removal, feature detection or license plate detection algorithms. In \cite{Wicaksono2017SpeedEO} and \cite{Tourani2019MotionbasedVS}, background subtraction is performed by means of the Gaussian Mixture Model. In \cite{9043732}, Scale Invariant Feature Transform (SIFT) and Speeded Up Robust Features (SURF) algorithms are exploited to detect feature points, whereas in \cite{Vakili}, the licence plate of the vehicle of interest is detected using an open source software and is tracked in consecutive frames to estimate the vehicle speed.

As a matter of fact, the literature on the extraction of information content from video signals is mainly based on heuristic ad-hoc solutions. To our knowledge, little or no attempts to employ sound approaches from estimation theory have been pursued and this paper wishes to contribute to fill this gap. As exceptions to this general trend, we mention previous contributions in which the ML criterion was successfully employed in the context of video processing for the extraction of periodic features \cite{7886029}, \cite{CATTANI2017158}. Despite a sound and consolidated method, the ML approach continues to be successfully applied to the solution of current problems. To mention a few, examples of recent works on ML parameters estimation include \cite{FASCISTA2021107907} and \cite{HAO2021108111}. In the former, a pseudo ML algorithm is developed to estimate the position of a multi-antenna receiver in dynamic multipath scenarios, whereas, in the latter, the position of multiple emitters is estimated. Another application of the ML approach can be found in \cite{WONG2021108087}, where a novel method for image reconstruction based on ML exposure level estimation is proposed.

Given a model that describes the observed data, the ML method allows to derive estimators of the unknown parameters of interest. In the application of interest in this paper, once the motion model of the video sequence is defined, the ML criterion can be applied to obtain an expression of the log-likelihood function to be optimized through a maximization process: the coordinates of the maximum of this function indicate the estimated components of the parameters of interest, namely the speed vector in this work. 

The performance of the proposed algorithm is compared with a more conventional solution, i.e. the block-matching method \cite[Ch. 4]{VideoProcessing}, that tends to be subject to several problems mainly related to the presence of noise, repeated patterns in the scene and the block size setting.
This paper expands upon preliminary conference contributions \cite{Eusipco}, \cite{TSP}.

The remainder of the paper is organized as follows. In Section~\ref{methodology}, preliminary video processing operations are described, along with the mathematical formulation of the dynamic motion model. In Section \ref{ml}, the likelihood function and the resulting speed estimation procedure are proposed. In Section~\ref{performance}, performance results on real videos are presented. Finally, in Section~\ref{conclusion} conclusions are drawn.

\section{Observation model} \label{methodology}

\subsection{Preliminary processing for object detection}\label{preprocessing}
\begin{figure*}[t!]
	\hspace*{-1.5cm}
	\centering
	\includegraphics[width=1.2\textwidth]{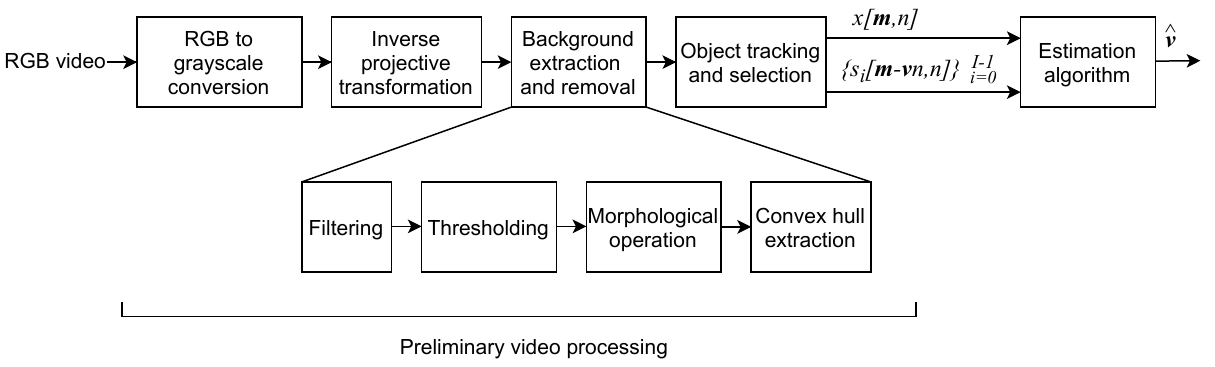}
	\caption{Overview of the proposed speed estimation method.}
	\label{fig1}
\end{figure*}

The extraction of foreground objects in a video sequence is at the basis of the proposed solution. An overview of the proposed method is shown in Figure \ref{fig1}.
In particular, an initial grayscale conversion from Red Green and Blue (RGB) input sequences is performed to obtain a grayscale video signal with frame period $T_{s}$. This conversion is motivated by the significant simplification it entails in the following processing operations and is highlighted in the first block of the diagram shown in Figure~\ref{fig1}. Frames are sampled at time instants $nT_{s}$, $n$ being the frame number, and can be described by matrices of $M_{1} \times M_{2}$ pixels. As already mentioned, in many realistic scenarios, foreground moving objects are likely to be subject to perspectival transformations arising from the projective mapping of 3D real world points to corresponding points on the 2D image plane~\cite{Solomon}. This mapping has also an immediate effect on the object speed: a constant speed in the real world may correspond to a non-constant speed in the image plane. However, it is possible to remove perspectival distortions by exploiting some image processing techniques that also allow to approximately recover the original shape and size of the objects of interest \cite{Solomon}, \cite{ViewGeom}. To this purpose, as highlighted in the second block of the diagram in Figure~\ref{fig1}, we may apply inverse projective transformations to each frame of the considered video sequence in order to compensate for the non-constant speed.

Once the original shape and speed of the framed objects is restored and can be considered constant in the 2D image plane, we proceed to detect them by removing the background, which is assumed to be static. The background removal operation is highlighted in the third block of the diagram in Figure~\ref{fig1} and is composed of four main steps, i.e., filtering, thresholding, morphological operation and convex hull extraction, as shown in the respective blocks in Figure~\ref{fig1}. Initially, we perform a basic image filtering operation based on the absolute difference of each frame and the background frame, i.e., reference frame, and we threshold the result to obtain a binary image where pixels belonging to the foreground have intensity equal to $1$ (white) and those belonging to the background have intensity equal to $0$ (black). When the considered video sequence is affected by noise, a spatial averaging filter can be applied to each frame to smooth the noise effect \cite{Solomon} and an estimate of the background can be obtained by a temporal average operation performed on the whole sequence. We then apply a cascade of morphological operations~\cite{Solomon}, i.e., erosion followed by dilation with different structuring elements, to reduce the misclassification of isolated objects and we extract the convex hull of the foreground objects to fill any remaining hole.

As a last step of this preliminary processing, object tracking and selection may be performed, as shown in Figure \ref{fig1}, based on standard feature extraction techniques \cite{Richard} to select a single object of interest, whereas the rest of the scene can be considered as background. Once a region of interest has been selected, a feature detection algorithm can be exploited to extract the locations of corners or other points of interest of the foreground object, i.e., a vehicle, at a specific frame. The locations of the detected features may be then searched in subsequent frames through feature tracking algorithms such as the Kanade-Lucas-Tomasi algorithm \cite{Tomasi91detectionand}.

The definitions of the notation $x[\mathbf{m},n]$, $\{s_{i}[\mathbf{m}- \mathbf{v}n,n]\}_{i=0}^{I-1}$ and $\hat{\mathbf{v}}$ in Figure~\ref{fig1} are provided in Sections \ref{model} and \ref{ml}.

\subsection{Dynamic motion model}\label{model}
Consider a frame size of $ M_{1} \times M_{2}$. The intensity of a pixel at position $\mathbf{m} = (m_{1}, m_{2})$ and frame $n$ can be defined as a discrete function $x[\mathbf{m},n]$, where $  0~\leq m_{1}~\leq M_{1} - 1 $ and $  0~\leq m_{2}~\leq M_{2} - 1 $.

Considering $ I $ framed objects subject to perspectival distortion, or any other dynamic change, the image of the $i$-th object can be denoted as $ s_{i}[\mathbf{m},n] $, with $ 0~\leq i \leq~I - 1 $. We can also consider a time dependent displacement vector $\boldsymbol{\delta}_{i}[n]=(\delta_{i,1}[n], \delta_{i,2}[n])^{\mathsf{T}}$, where $ \delta_{i,1}[n] $ and $ \delta_{i,2}[n] $ represent the horizontal and vertical components, respectively. Following the approach in \cite{Bri2}, and defining $ s_{i}[\mathbf{m} - \boldsymbol{\delta}_{i}[n], n] $ as a shift that affects the $i$-th object in the 2D image plane at the $n$-th frame, we can model the pixel intensities of the grayscale video signal as
\begin{equation}
\begin{aligned}
x[\mathbf{m},n] = \; b[\mathbf{m}] + \sum_{i=0}^{I-1} s_{i}[\mathbf{m} - \boldsymbol{\delta}_{i}[n], n] + v_{b}[\mathbf{m},n] + w[\mathbf{m},n]
\label{eq:model1}
\end{aligned}
\end{equation}
where $ b[\mathbf{m}] $ is the static background, whose partial occlusion/un-occlusion due to the motion of the objects is taken into account by the term $ v_{b}[\mathbf{m},n] $, and $ w[\mathbf{m},n] $ represents samples of independent and identically distributed (i.i.d.) zero-mean Gaussian noise. When inverse projective transformations are applied and the shape and size of the framed objects can be considered constant, their images and shifts in the 2D image plane can be simplified as $ s_{i}[\mathbf{m}] $ and $ s_{i}[\mathbf{m} - \boldsymbol{\delta}_{i}[n]] $, respectively.

The $I$ objects are now assumed to be moving with comparable and almost constant speed. This assumption is a basic requirement for any speed estimation problem, as the speed must be almost constant over a window of consecutive frames of sufficient duration in order to enable its correct estimation. Hence, we can express the common displacement term as  $ \boldsymbol{\delta}[n] = \mathbf{v} n$, where $ \mathbf{v} = (v_{1}, v_{2})^{\mathsf{T}}$ is the vector of the common uniform speed, measured in pixel/frame, to be estimated. Accordingly, the model in (\ref{eq:model1}) can be written~as
\begin{equation}
\begin{aligned}
x[\mathbf{m},n] =  \; b[\mathbf{m}] + \sum_{i=0}^{I-1} s_{i}[\mathbf{m} -\mathbf{v} n, n] + v_{b}[\mathbf{m},n] + w[\mathbf{m},n].
\label{eq:model2}
\end{aligned}
\end{equation}

By the preliminary background removal operation discussed in Section \ref{preprocessing} and corresponding to the third block of the diagram shown in Figure \ref{fig1}, it is possible to further simplify the observation model in (\ref{eq:model2}) as
\begin{equation}
x[\mathbf{m},n] = \sum_{i=0}^{I-1} s_{i}[\mathbf{m} -\mathbf{v} n, n] + w[\mathbf{m},n]
\label{eq:model_nobk}
\end{equation}
where the background-related terms have been neglected and the observation sequence $x[\mathbf{m},n]$ is obtained after the background removal operation.

Assume first that $ \mathbf{v} $ has integer components in pixel/frame. Further processing can be conveniently performed in the Fourier Transform (FT) domain, which provides the advantage of expressing a displacement as a linear phase term thanks to the shift theorem, as also discussed in \cite{Bri2}. Hence, applying the definition of the Discrete Fourier Transform (DFT) of a generic 2D discrete function \cite{Kay}, we can express the frequency domain equivalent of the model in (\ref{eq:model_nobk}) as
\begin{equation}
\begin{aligned}
X[\mathbf{k},n]  = \sum_{i=0}^{I-1} S_{i}[\mathbf{k}, n]e^{-j2\pi\mathbf{u_{k}}^{\mathsf{T}}\mathbf{v}n}  + W[\mathbf{k},n]
\label{eq:freqmodel1}
\end{aligned}
\end{equation}
where $\mathbf{k}~=~(k_{1},k_{2})^\mathsf{T} $ is the vector of the two discrete indices of the 2D DFT, with $ 0 \leq k_{l} \leq M_{l} -1 $, $ l ~=~1,2 $, $ \mathbf{u_{k}}~=~\left( k_{1}/M_{1} , k_{2}/M_{2} \right) ^\mathsf{T} $ is the vector of the normalized spatial frequencies and uppercase letters denote the DFTs of the corresponding signals in (\ref{eq:model_nobk}).

Consider now the case of a fractional value of the speed vector $\mathbf{v}$, denote the number of sub-pixel quantization levels by the integer $F$ and assume $\mathbf{v}$ is quantized accordingly. The displacement vector can be written as 
\begin{equation}
\mathbf{v} n=\mathbf{d}[\mathbf{v},n] + \frac{\mathbf{f}[\mathbf{v},n]}{F}=\left(d_{1}[v_{1},n]+\frac{f_{1}[v_{1},n]}{F}, d_{2}[v_{2},n]+\frac{f_{2}[v_{2},n]}{F}\right)^{\mathsf{T}}
\label{eq:fracdisp}
\end{equation}
where
\begin{align}
\label{eq_d} \mathbf{d}[\mathbf{v},n] & =(d_{1}[v_{1},n], d_{2}[v_{2},n])^{\mathsf{T}}= \Big \lfloor \mathbf{v} n \Big \rfloor \\
\label{eq_f} \frac{\mathbf{f}[\mathbf{v},n]}{F} & =\frac{\left(f_{1}[v_{1},n], f_{2}[v_{2},n]\right)^{\mathsf{T}}}{F}= \{\mathbf{v} n \}
\end{align}
represent the integer and fractional parts of the vector, respectively, with $  f_{i}[v_{i}, n] \in \{0, 1, 2, \ldots, F-1 \} $, $i = 1,2$, $\big \lfloor \cdot \big \rfloor$ denotes the floor function and $ \{x \}~=~x~-~\big \lfloor x \big \rfloor $.

In order to extend the model in (\ref{eq:model_nobk}) to the most general case where a foreground object may shift with a fractional speed, it is useful to define the fractional sub-pixel translation $y[\mathbf{m}]$ of an image $s[\mathbf{m}]$ with fractional displacement $ \left( \frac{f_{1}}{F}, \frac{f_{2}}{F} \right) $, $f_{i} = 0, 1, 2, \ldots, F-1 $, $i = 1,2$, as
\begin{equation}
\begin{aligned}
y[\mathbf{m}] = & \bigg(1- \frac{f_{1}}{F}\bigg)\bigg(1- \frac{f_{2}}{F}\bigg)s[\mathbf{m}] \\& +\frac{f_{1}}{F}\bigg(1- \frac{f_{2}}{F}\bigg)s\big[\mathbf{m} - \mathbf{h}_{1}\big] \\& + \bigg(1- \frac{f_{1}}{F}\bigg)\frac{f_{2}}{F}s\big[\mathbf{m} - \mathbf{h}_{2}\big] \\& + \frac{f_{1}}{F} \frac{f_{2}}{F} s\big[\mathbf{m} - \mathbf{h}_{1} - \mathbf{h}_{2}\big]
\label{eq:image_translation}
\end{aligned}
\end{equation}
where $\mathbf{h}_{1} = (1, 0)^{\mathsf{T}}$ and $\mathbf{h}_{2} = (0, 1)^{\mathsf{T}}$ are the unitary vectors related to the two components. The model in (\ref{eq:model_nobk}) can be thus expanded as:
\begin{eqnarray}
\begin{aligned}
x[\mathbf{m},n] = & \bigg(1- \frac{f_{1}[v_{1},n]}{F}\bigg)\bigg(1- \frac{f_{2}[v_{2},n]}{F}\bigg)\sum_{i=0}^{I-1} s_{i}\big[\mathbf{m}- \mathbf{d}[\mathbf{v},n],n\big] \\& +\frac{f_{1}[v_{1},n]}{F}\bigg(1- \frac{f_{2}[v_{2},n]}{F}\bigg)\sum_{i=0}^{I-1} s_{i}\big[\mathbf{m} - \mathbf{d}[\mathbf{v},n]-\mathbf{h}_{1},n\big] \\& + \bigg(1- \frac{f_{1}[v_{1},n]}{F}\bigg)\frac{f_{2}[v_{2},n]}{F}\sum_{i=0}^{I-1} s_{i}\big[\mathbf{m} - \mathbf{d}[\mathbf{v},n]- \mathbf{h}_{2},n\big] \\& +\frac{f_{1}[v_{1},n]}{F} \frac{f_{2}[v_{2},n]}{F} \sum_{i=0}^{I-1} s_{i}\big[\mathbf{m} - \mathbf{d}[\mathbf{v},n]- \mathbf{h}_{1} - \mathbf{h}_{2},n\big] + w[\mathbf{m},n].
\label{eq:new_model}
\end{aligned}
\end{eqnarray}


Taking the 2D DFT of (\ref{eq:new_model}), an equivalent observation model in the frequency domain can be obtained. Using again the shift theorem, this model can be formulated as
\begin{equation}
\begin{aligned}
X[\mathbf{k},n]  = \sum_{i=0}^{I-1} S_{i}[\mathbf{k}, n]e^{-j2\pi\mathbf{u_{k}}^{\mathsf{T}}\mathbf{d}[\mathbf{v},n]}a[\mathbf{v},n]  + W[\mathbf{k},n]
\end{aligned}
\label{eq:new_model_f}
\end{equation}
where
\begin{equation}
\begin{aligned}
a[\mathbf{v},n]=& \bigg(1- \frac{f_{1}[v_{1},n]}{F}\bigg)\bigg(1- \frac{f_{2}[v_{2},n]}{F}\bigg) \\&
+\frac{f_{1}[v_{1},n]}{F}\bigg(1- \frac{f_{2}[v_{2},n]}{F}\bigg) e^{-j2\pi\frac{k_{1}}{M_{1}}} \\&
+\bigg(1- \frac{f_{1}[v_{1},n]}{F}\bigg)\frac{f_{2}[v_{2},n]}{F} e^{-j2\pi\frac{k_{2}}{M_{2}}} \\&
+ \frac{f_{1}[v_{1},n]}{F} \frac{f_{2}[v_{2},n]}{F} e^{-j2\pi\big(\frac{k_{1}}{M_{1}}+\frac{k_{2}}{M_{2}}\big)}.
\label{a_n}
\end{aligned}
\end{equation}

\section{Maximum likelihood speed estimation}\label{ml}
Observing now that the model in (\ref{eq:new_model_f}) describes Gaussian observations that are independent in the spatial and discrete frequency domains, ML estimation can be used to derive an expression of the estimator $\hat{\mathbf{v}}$ of the unknown speed vector \cite{Kay}. The dependence of (\ref{eq:new_model_f}) on the speed vector is through the terms $ \mathbf{d}[\mathbf{v},n] $ and $ a[\mathbf{v},n] $. 

Considering an observation window of $N$ frames, the relevant likelihood function of the model in (\ref{eq:new_model_f}) is
\begin{equation}
\begin{aligned}
p\big( X[\mathbf{k},0 ]&\cdots X[\mathbf{k},N-1 ];\mathbf{v}\big) \\ = & \left( \frac{1}{2\pi\sigma^{2}}\right)^\frac{M_{1}M_{2}N}{2}  \cdot \exp \Bigg\{  -\frac{1}{2\sigma^2}\sum_{k_{1}=0}^{M_{1}-1}\sum_{k_{2}=0}^{M_{2}-1}\sum_{n=0}^{N-1} \Bigg| X[\mathbf{k},n] \\& - \sum_{i=0}^{I-1} S_{i}[\mathbf{k},n]e^{-j2\pi\mathbf{u_{k}}^{\mathsf{T}}\mathbf{d}[\mathbf{v},n]} a[\mathbf{v},n] \Bigg|^2 \Bigg\}
\end{aligned}
\label{likelihood}
\end{equation}
where $\sigma$ is the standard deviation of the additive Gaussian noise elements.

We can also derive the log-likelihood function from (\ref{likelihood}) as
\begin{equation}
\begin{aligned}
\ln \big( p&(X[\mathbf{k},0 ]\cdots X[\mathbf{k},N-1 ];\mathbf{v})\big) = \underbrace{ - \frac{M_{1}M_{2}N}{2} \ln(2\pi\sigma^{2})}_{\text{(a)}} \\& - \frac{1}{2\sigma^2} \underbrace{\sum_{k_{1}=0}^{M_{1}-1}\sum_{k_{2}=0}^{M_{2}-1}\sum_{n=0}^{N-1} \left| X[\mathbf{k},n] - \sum_{i=0}^{I-1} S_{i}[\mathbf{k}]e^{-j2\pi\mathbf{u_{k}}^{\mathsf{T}}\mathbf{d}[\mathbf{v},n]}a[\mathbf{v},n] \right|^2}_{\text{(b)}}
\label{eq:logLF_bck}
\end{aligned}
\end{equation}
where some terms are highlighted. In particular, given that (a) is a constant term and the multiplicative coefficient $- \frac{1}{2\sigma^2}$ is a constant factor, in the sense that they do not depend on the trial speed value $ \mathbf{v} $, they are irrelevant for the estimation problem and can be discarded. The term (b) can be equivalently minimized with respect to the value of $\mathbf{v}$. Hence, an equivalent likelihood function to be minimized is
\begin{eqnarray}
\sum_{k_{1}=0}^{M_{1}-1}\sum_{k_{2}=0}^{M_{2}-1}\sum_{n=0}^{N-1} \left| X[\mathbf{k},n] - \sum_{i=0}^{I-1} S_{i}[\mathbf{k}]e^{-j2\pi\mathbf{u_{k}}^{\mathsf{T}}\mathbf{d}[\mathbf{v},n]}a[\mathbf{v},n] \right|^2
\label{eq:logLF_bck2}.
\end{eqnarray}

We can explicitly express (\ref{eq:logLF_bck2}) as
\begin{equation}
\begin{aligned} 
\sum_{k_{1}=0}^{M_{1}-1} & \sum_{k_{2}=0}^{M_{2}-1} \sum_{n=0}^{N-1}   \bigg\{ \left| X[\mathbf{k},n] \right|^{2} + \bigg|\sum_{i=0}^{I-1} S_{i}[\mathbf{k},n]a[\mathbf{v},n] \bigg|^{2} \\&  - 2 \sum_{i=0}^{I-1}  \operatorname{Re} \big\{ X[\mathbf{k},n] S_{i}^{\ast}[\mathbf{k},n]e^{j2\pi\mathbf{u_{k}}^{\mathsf{T}}\mathbf{d}[\mathbf{v},n]}a^{\ast}[\mathbf{v},n] \big\}\bigg\}
\end{aligned}
\label{eq:term_a_b}
\end{equation} 
where $\operatorname{Re} \{\cdot\}$ and $ \left( \cdot \right) ^{\ast}$ are the real part and the complex conjugate operators, respectively. The quadratic terms in (\ref{eq:term_a_b}) are irrelevant or practically so. In fact, the term $| X[\mathbf{k},n] |^{2}$ is independent of $\mathbf{v}$ and is irrelevant. The term $\left|\sum_{i=0}^{I-1} S_{i}[\mathbf{k},n]a[\mathbf{v},n] \right|^{2}$ depends on $\mathbf{v}$ through the factor $a[\mathbf{v},n]$, that depends only on the fractional part of the speed vector. In particular, if a grid search with a resolution of $\frac{1}{F}$ pixel/frame is implemented, then $F^{2}$ possible values of $a[\mathbf{v},n]$ are obtained according to (\ref{a_n}), which repeat periodically over the grid of possible values. It turns out that these values are subject to very small variations if compared against the mixed term. As consequence, the term $\left|\sum_{i=0}^{I-1} S_{i}[\mathbf{k},n]a[\mathbf{v},n] \right|^{2}$ can be considered almost constant, hence practically also irrelevant, and (\ref{eq:term_a_b}) can be minimized by maximizing the following approximate likelihood function
\begin{equation}
\begin{aligned} 
\sum_{k_{1}=0}^{M_{1}-1} \sum_{k_{2}=0}^{M_{2}-1} \sum_{n=0}^{N-1} \operatorname{Re}\bigg\{  \sum_{i=0}^{I-1}  X[\mathbf{k},n] S_{i}^{\ast}[\mathbf{k},n]e^{j2\pi\mathbf{u_{k}}^{\mathsf{T}}\mathbf{d}[\mathbf{v},n]}a^{\ast}[\mathbf{v},n] \bigg\}
\end{aligned}
\label{eq:term_a_b2}
\end{equation}
where the linearity of $\operatorname{Re} \{\cdot\}$ and sum operators has been exploited. The accuracy of this approximate likelihood function will be discussed and numerically demonstrated in the next section.

Finally, the following expression for the (quasi) ML speed estimator is obtained:
\begin{equation}
\begin{aligned}
\hat{\mathbf{v}} = \argmax_{\mathbf{v}} \sum_{k_{1}=0}^{M_{1}-1} \sum_{k_{2}=0}^{M_{2}-1} \sum_{n=0}^{N-1}  \operatorname{Re} \bigg\{  \sum_{i=0}^{I-1} X[\mathbf{k},n]  \cdot S_{i}^{\ast}[\mathbf{k},n]e^{j2\pi\mathbf{u_{k}}^{\mathsf{T}}\mathbf{d}[\mathbf{v},n]}a^{\ast}[\mathbf{v},n] \bigg\}.
\end{aligned}
\label{eq_vhat_nobk}
\end{equation}

In the case of integer values of displacement components, $\mathbf{d}[\mathbf{v},n]~=~\mathbf{v} n$ and $a[\mathbf{v},n] = 1$ in (\ref{a_n})-(\ref{eq_vhat_nobk}), thus a simplified version of the proposed solution is obtained that corresponds to the one considered in \cite{Eusipco} and \cite{TSP}. In this particular case, it is possible to express (\ref{eq_vhat_nobk}) in the following compact form:
\begin{equation}
\begin{aligned}
\hat{\mathbf{v}} = \argmax_{\mathbf{v}} \sum_{k_{1}=0}^{M_{1}-1} \sum_{k_{2}=0}^{M_{2}-1} \operatorname{Re} \Bigg\{  Y \bigg[\mathbf{k}, -\frac{\mathbf{u_{k}}^{\mathsf{T}}\mathbf{v}}{T_{s}} \bigg] \Bigg\}
\end{aligned}
\label{eq_vhat}
\end{equation}
where
\begin{equation}
\begin{aligned}
&Y \left[\mathbf{k}, q\right] = \sum_{n=0}^{N-1}  \sum_{i=0}^{I-1} X[\mathbf{k},n] S_{i}^{\ast}[\mathbf{k},n]e^{-j2 \pi q T_{s} n}
\end{aligned}
\label{eq:ft_nobck}
\end{equation}
is the continuous-frequency FT of the temporal sequence $ \bigl\{ \sum_{i=0}^{I-1} X[\mathbf{k},n] S_{i}^{\ast}[\mathbf{k},n] \bigr\}  $ in the continuous-frequency variable $q$. The function (\ref{eq:ft_nobck}) can be used with $ q = -\frac{\mathbf{u_{k}}^{\mathsf{T}}\mathbf{v}}{T_{s}}$ to obtain (\ref{eq_vhat}).\\

The estimated speed vector  $\hat{\mathbf{v}}$ is specified by the coordinates of the maximum of the log-likelihood function defined in agreement with (\ref{eq_vhat_nobk}) as follows
\begin{equation}
\begin{aligned}
J(\mathbf{v}) = \frac{1}{M_{1} M_{2}} \sum_{k_{1}=0}^{M_{1}-1} \sum_{k_{2}=0}^{M_{2}-1}\sum_{n=0}^{N-1} \operatorname{Re} \Bigg\{  \sum_{i=0}^{I-1} X[\mathbf{k},n]  \cdot S_{i}^{\ast}[\mathbf{k},n]e^{j2\pi\mathbf{u_{k}}^{\mathsf{T}}\mathbf{d}[\mathbf{v},n]}a^{\ast}[\mathbf{v},n] \Bigg\}
\end{aligned}
\label{eq_J}
\end{equation}
where the normalization coefficient $ 1/M_{1} M_{2} $ is introduced to reduce the dynamic range of the log-likelihood function by several orders of magnitude without any impact on the final estimate.

As described in Section \ref{model}, inverse projective transformations can be applied to recover the original shape and size of the objects of interest, whose images can thus be considered constant over time. If the image of the objects of interest can be considered constant with respect to $n$, the definition in (\ref{eq_J}) can be simplified as follows:
\begin{equation}
\begin{aligned}
J(\mathbf{v}) = \frac{1}{M_{1} M_{2}} \sum_{k_{1}=0}^{M_{1}-1} \sum_{k_{2}=0}^{M_{2}-1}\sum_{n=0}^{N-1} \operatorname{Re} \Bigg\{  \sum_{i=0}^{I-1} X[\mathbf{k},n]  \cdot S_{i}^{\ast}[\mathbf{k}]e^{j2\pi\mathbf{u_{k}}^{\mathsf{T}}\mathbf{d}[\mathbf{v},n]}a^{\ast}[\mathbf{v},n] \Bigg\}.
\end{aligned}
\label{eq_J2}
\end{equation}

\section{Applications and results} \label{performance}
In this section, we discuss the performance of the proposed algorithm on the basis of some experimental results directly obtained by maximizing (\ref{eq_J}) or (\ref{eq_J2}). In particular, since reasonable assumptions about the range of values of the correct speed can be made, we can implement a simple grid search to find the optimal value of $\hat{\mathbf{v}}$. Iterative gradient-search approaches could also be considered to expedite the numerical solution~\cite{Kay}, but are not pursued here because out of the scope of the present paper.

\begin{figure*}[t!]
	\centering
	\includegraphics[width=0.85\textwidth]{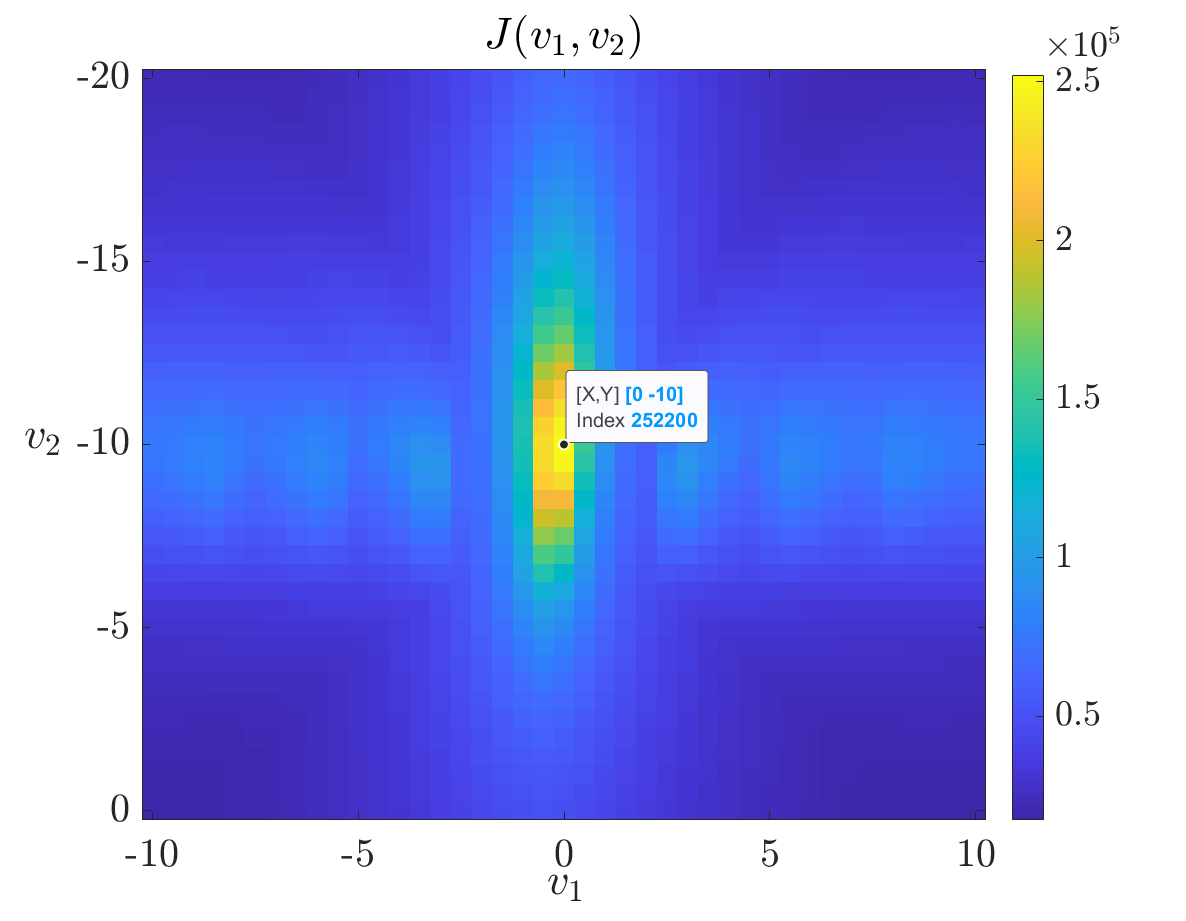}
	\caption{Example of the log-likelihood function (\ref{eq_J}) plotted versus speed components.}
	\label{J_function}
\end{figure*}

As an illustrative example, the log-likelihood function in (\ref{eq_J}), for one of the studied cases, is displayed in Figure \ref{J_function} versus the components of the speed vector $\mathbf{v}$ with a grid resolution of 0.5 pixel/frame for both components, (i.e., $F = 2$ in (\ref{a_n})). The peak of the function, whose coordinates indicate the estimated speed value, is highlighted.

For the sake of completeness, the log-likelihood function in Figure \ref{J_function} is now compared with the quadratic term in (\ref{eq:term_a_b}), that has been neglected to derive the approximate likelihood function (\ref{eq_J}). Accounting for the proper scaling factor $ 1/2 $, this term is:
\begin{equation}
\begin{aligned} 
\gamma(\mathbf{v}) = \sum_{k_{1}=0}^{M_{1}-1} \sum_{k_{2}=0}^{M_{2}-1} \sum_{n=0}^{N-1} \frac{1}{2}  \bigg|\sum_{i=0}^{I-1} S_{i}[\mathbf{k},n]a[\mathbf{v},n] \bigg|^{2} .
\end{aligned}
\label{eq:gamma}
\end{equation}
\begin{figure*}[t!]
	\hspace*{-1.5cm}
	\begin{subfigure}{0.6\textwidth}
		\includegraphics[width=1\textwidth]{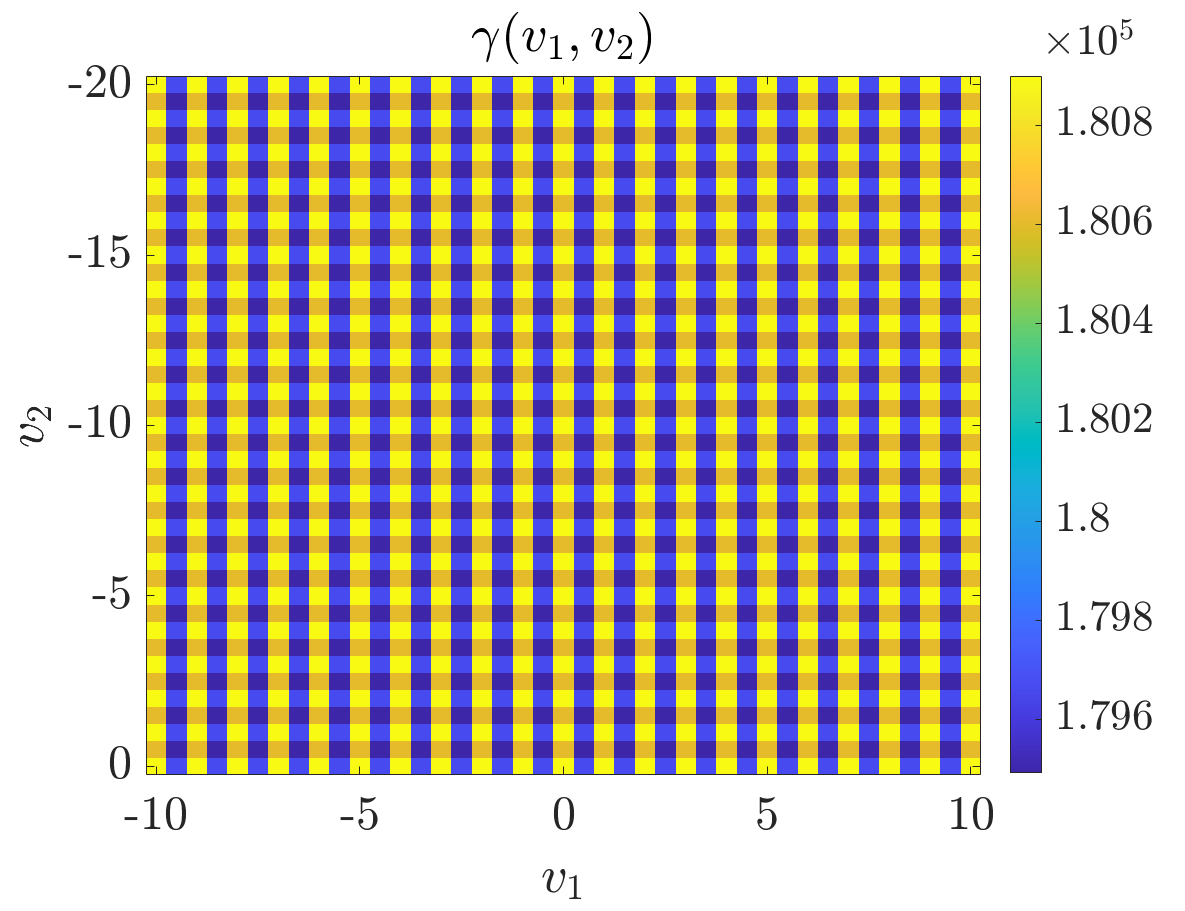}
		\caption{}
		\label{gamma}
	\end{subfigure}
	\begin{subfigure}{0.6\textwidth}
		\includegraphics[width=1\textwidth]{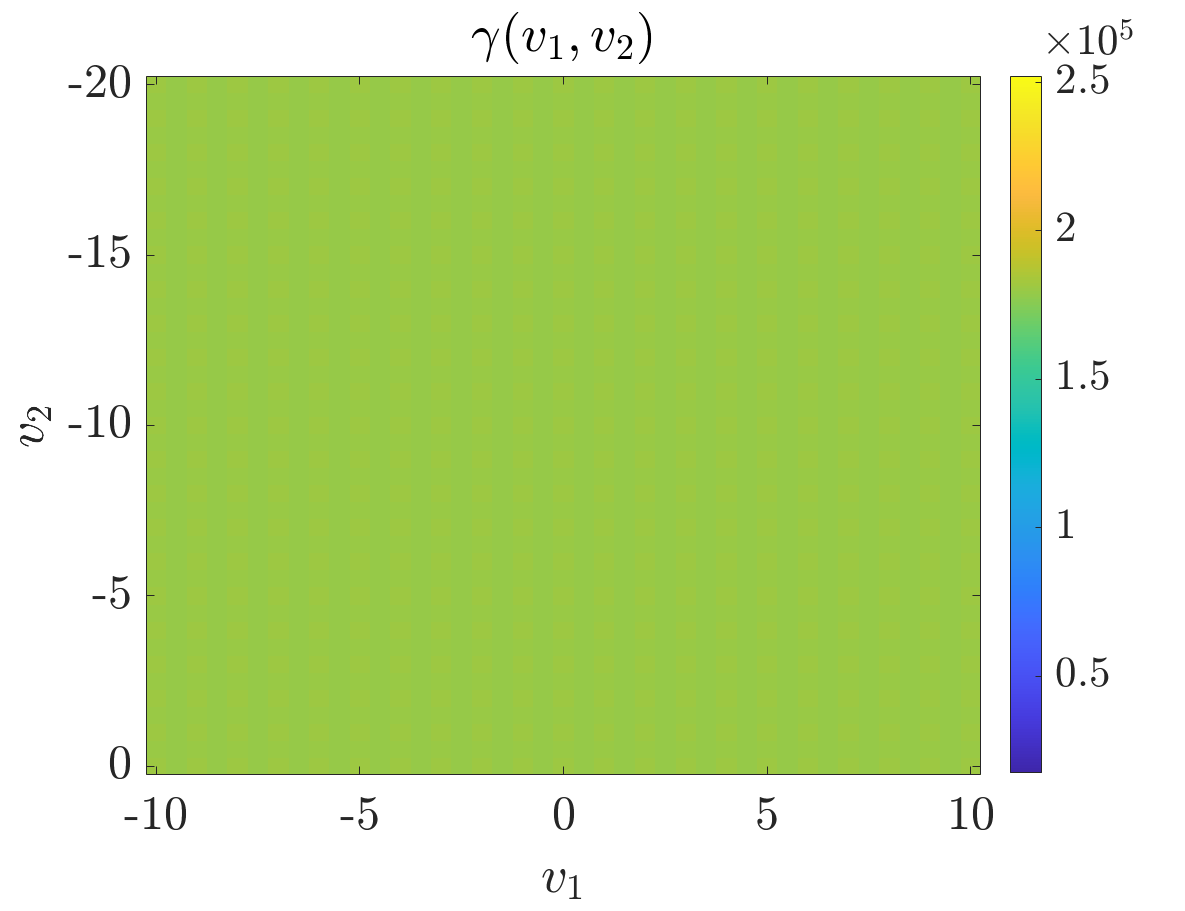}
		\caption{}
		\label{gamma2}
	\end{subfigure}
	\caption{Term in (\ref{eq:gamma}) plotted versus speed components for two different visualization scales.}
	\label{gam}
\end{figure*} 
The term in (\ref{eq:gamma}), obtained in agreement with the log-likelihood function shown in Figure~\ref{J_function}, is plotted in Figure~\ref{gam}, where two different scales are considered for the sake of visualization. In particular, the scale used in Figure~\ref{gam}(a) allows to highlight the small variations of this term. The scale used in Figure~\ref{gam}(b) is equal to the one in Figure~\ref{J_function} and allows to highlight the practically constant behaviour of this term with respect to the log-likelihood function in Figure \ref{J_function}. This comparison demonstrates the accuracy of the presented approximate ML estimator. Note that, for the considered resolution of 0.5 pixel/frame, only four values of $a[\mathbf{v},n]$ and $ \gamma(\mathbf{v}) $ are obtained in expressions (\ref{a_n}) and (\ref{eq:gamma}). The four values in $ \gamma(\mathbf{v}) $ are clearly visible in Figure~\ref{gam}(a).


\subsection{Performance measure}
As the operations of object tracking and selection described in Section \ref{intro} can be exploited to focus on a single object, for the sake of simplicity, scenes framing a single moving object are considered in this section, i.e., $I~=~1$ in (\ref{eq_J}) and (\ref{eq_J2}). A few sets of real videos were recorded with different camera angles. The performance of the estimation algorithm is analysed in terms of Root Mean Square (RMS) Error (RMSE) between the estimated speed vector and the correct speed, which is manually measured. The proposed algorithm is also tested in the presence of noise to assess its robustness. To this end, Gaussian noise with spatially and temporally i.i.d. elements is superimposed to each video sequence. To smooth the noise effect, a spatially averaging filter with size $ 7 \times 7 $ pixel is applied.
Considering $ R $ different noise realizations and $ J $ analysed videos, we define the RMSE normalized to the RMS value of the correct speed by averaging over the noise realizations and video sequences as
\begin{equation}
\begin{aligned}
\varepsilon = \sqrt{\frac{\sum\limits_{r=1}^{R} \sum\limits_{j=1}^{J}\big|\hat{\mathbf{v}}_{r,j}-\mathbf{v}_{j}\big|^{2}}{R \sum\limits_{j=1}^{J} |\mathbf{v}_{j}|^{2}}}
\end{aligned}
\label{eq_rmse}
\end{equation}
where $\mathbf{v}_{j}$ and $\hat{\mathbf{v}}_{r,j}$ are the correct speed components for the $j$-th video and the estimated speed components for the $ j $-th video and the $ r $-th noise realization, respectively.

The obtained results are compared with the performance of the block-matching method \cite[Ch. 4]{VideoProcessing}. At first, the normalized RMSE in (\ref{eq_rmse}) is investigated for increasing values of the noise variance $ \sigma^2 $ for single video sequences and a small set of video sequences in which the same camera viewpoint, position and location are preserved. The overall performance is finally evaluated on all considered scenarios for increasing values of the peak signal to average noise power ratio, or Signal to Noise Ratio (SNR) for brevity, hence for decreasing values of the noise variance $ \sigma^2 $.

\subsection{Speed estimation}
The proposed estimation algorithm is tested for 10 noise realizations on 12 sequences extracted from 5 real videos specifically recorded. Various camera angles and locations are considered in order to assess the robustness of the proposed method in different perspectival conditions. The preprocessing operations described in Section \ref{preprocessing} need to be calibrated for each set of videos recorded with the same camera setting, position and location. In particular, the inverse projective transformation and dimension of the structuring elements of the morphological operation need to be properly set. As multiple cars are captured in some of the recorded videos, it is sufficient to trim and crop the sequences to focus on a single vehicle at a time. The duration of the trimmed video sequences is variable and ranges from 25 to 165 frames. Also, the size and the frame rate depend on the setting of the employed recording device. In particular, here the frame rate is set as $f_{s} = 25$ or 30 Hz, whereas the frame size for all videos is converted to a fixed size of $800 \times 300$ or $300 \times 800$ pixel after the inverse projective transformation. 

\begin{figure*}[t!]
	\centering
	\begin{tabular}{cccc}
		frame 1 & frame 81 & frame 151\\
		\includegraphics[width = 3.1cm, height=4.5cm]{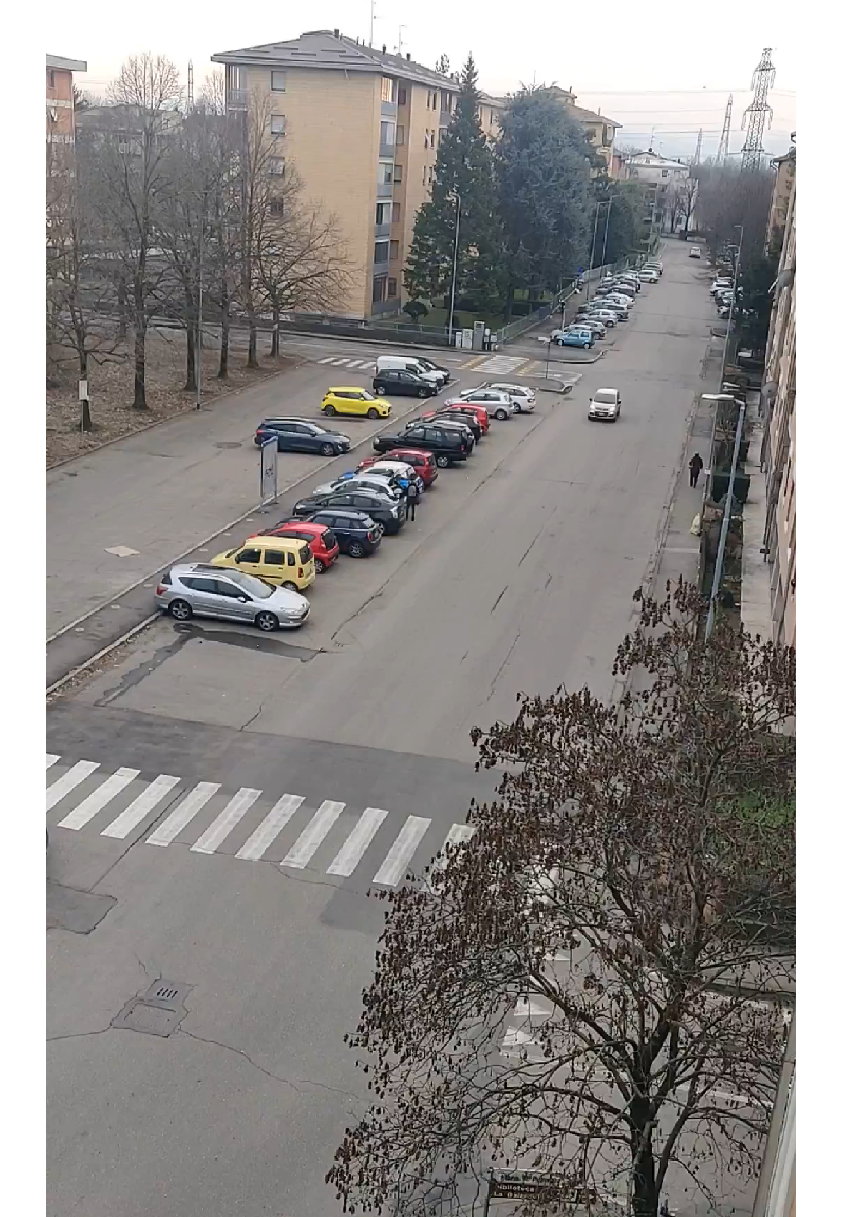} &
		\includegraphics[width = 3.1cm, height=4.5cm]{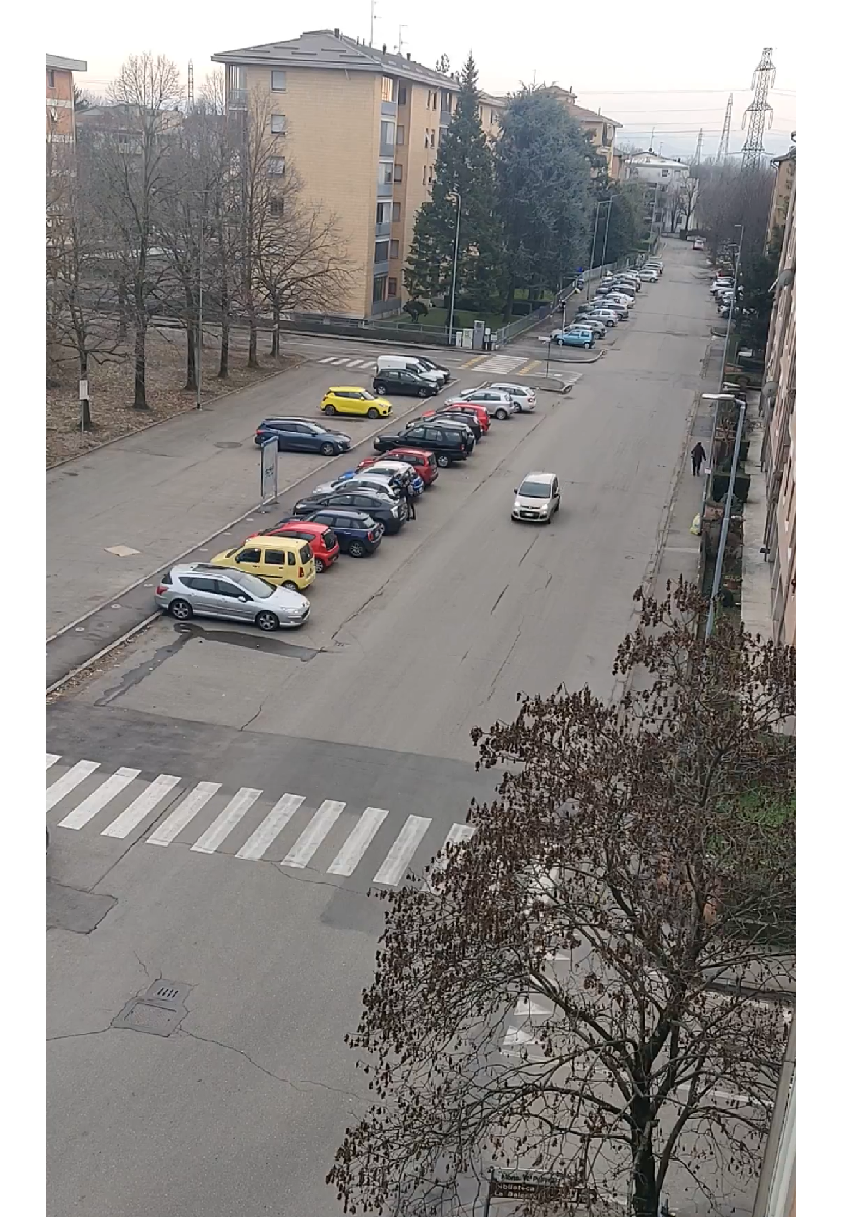} &
		\includegraphics[width = 3.1cm, height=4.5cm]{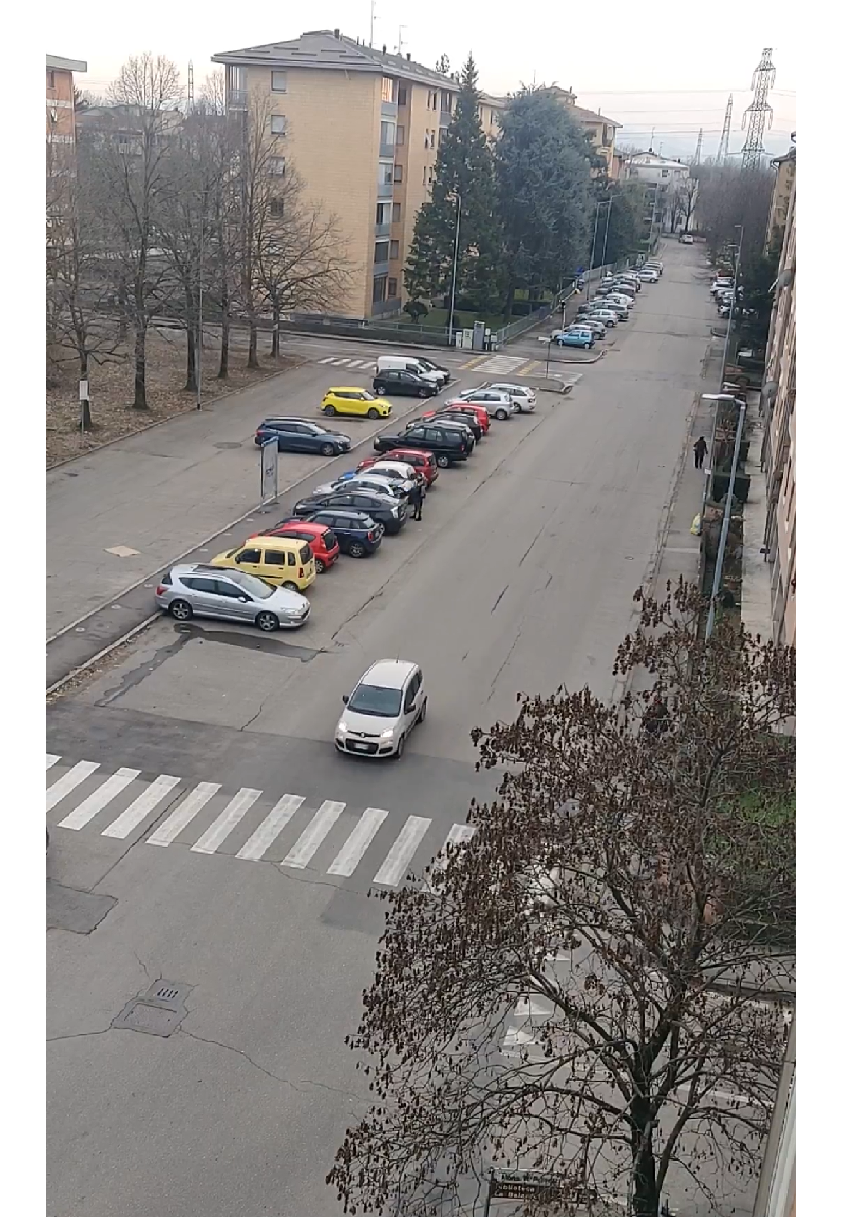} &
		(a)\\
		\includegraphics[width = 2.8cm, height=4.5cm]{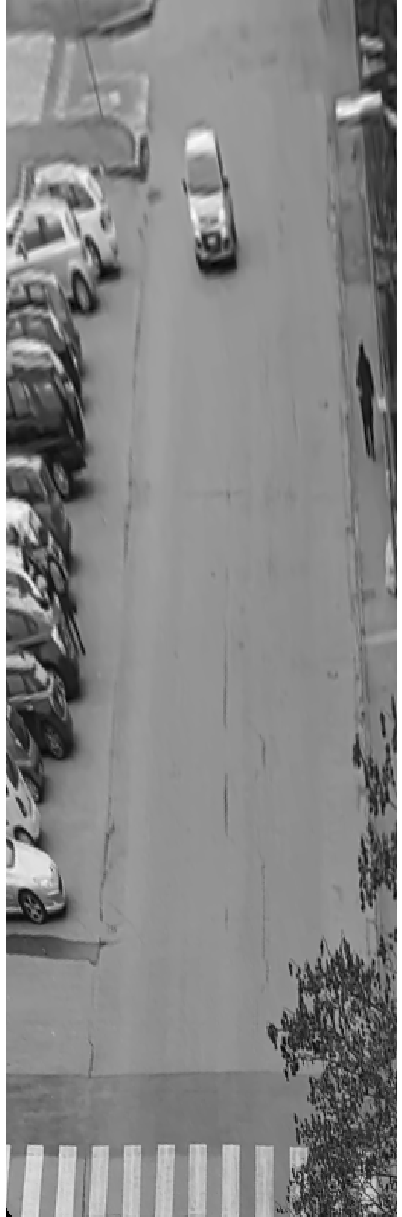} &
		\includegraphics[width = 2.8cm, height=4.5cm]{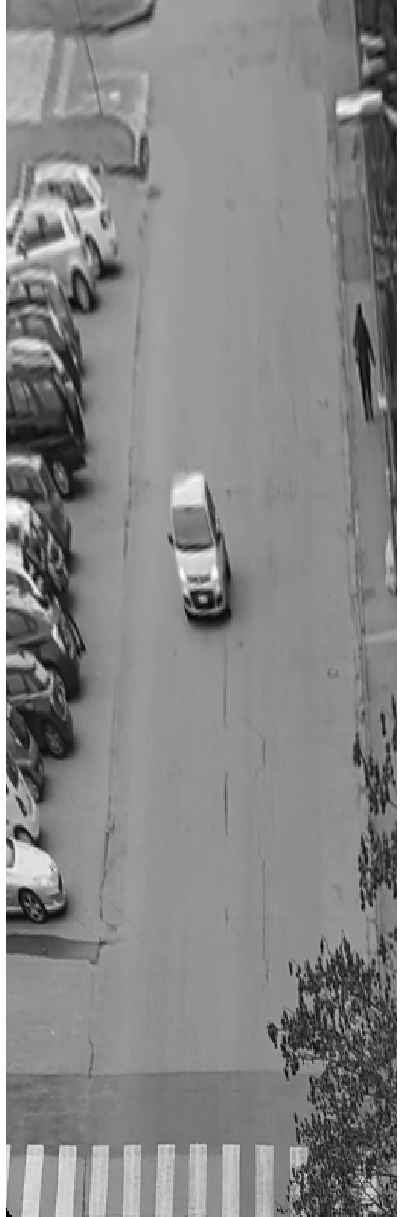} &
		\includegraphics[width = 2.8cm, height=4.5cm]{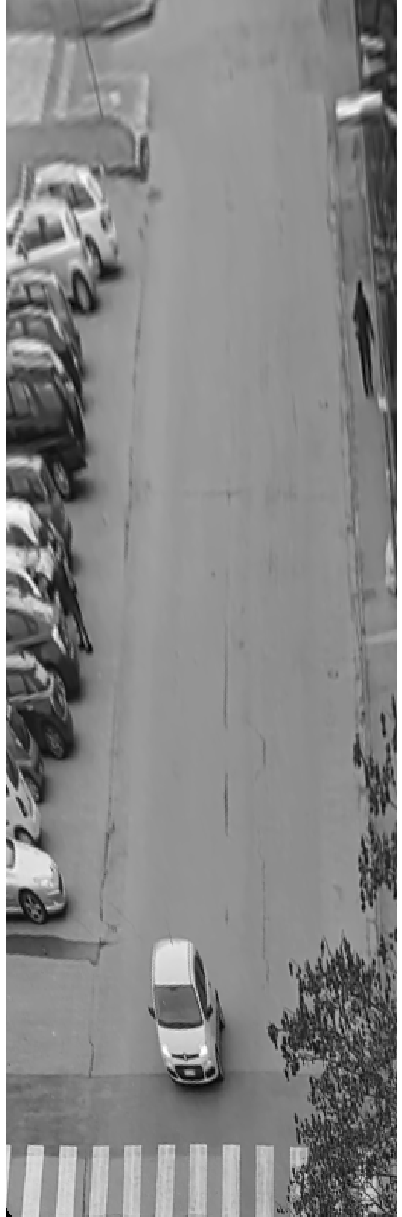} &
		(b)\\
		\includegraphics[width = 2.8cm, height=4.5cm]{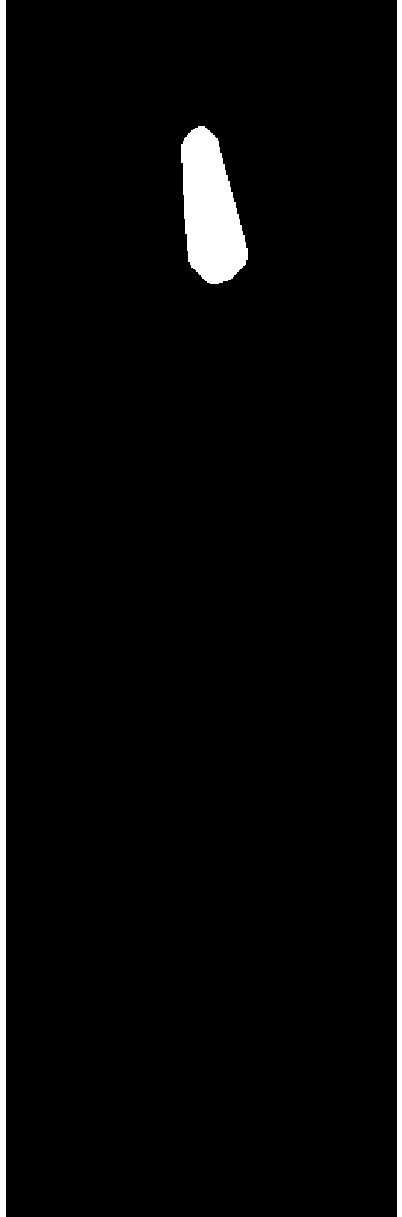} &
		\includegraphics[width = 2.8cm, height=4.5cm]{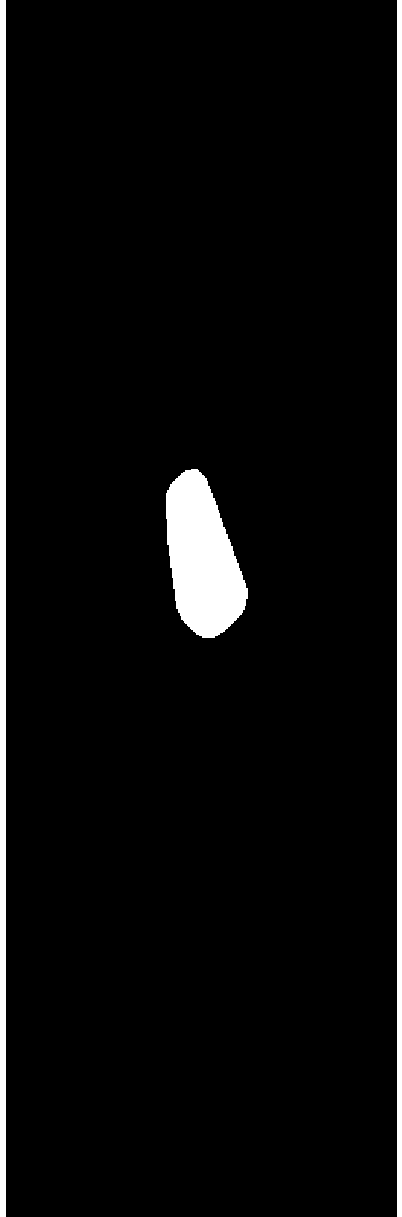} &
		\includegraphics[width = 2.8cm, height=4.5cm]{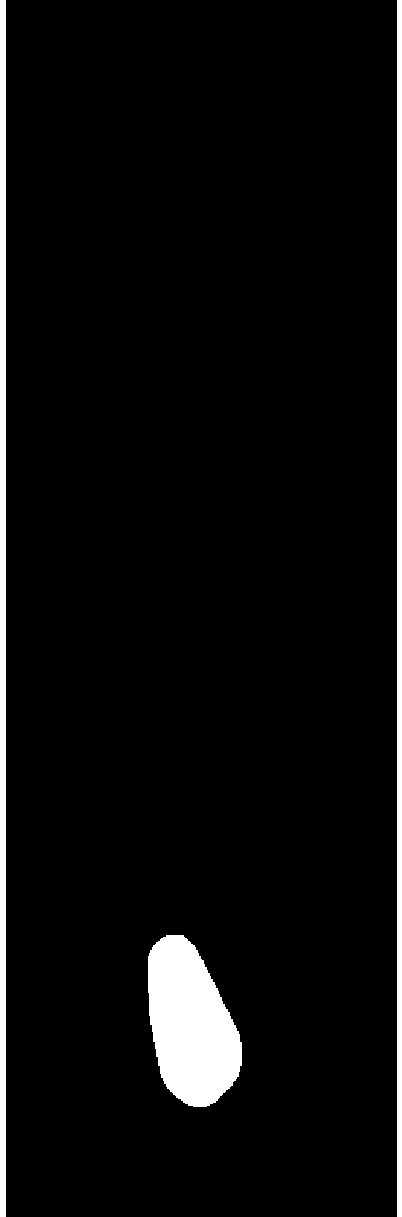} &
		(c)
	\end{tabular}
	\caption{Sample frames of: (a) original Sequence 1, (b) processed sequence after the inverse projective transformation, (c) processed sequence after background removal and convex hull extraction.}
	\label{frame}
\end{figure*}

\begin{figure*}[t!]
	\centering
	\begin{tabular}{cccc}
		frame 91 & frame 121 & frame 161\\
		\includegraphics[width = 3.9cm, height=2.4cm]{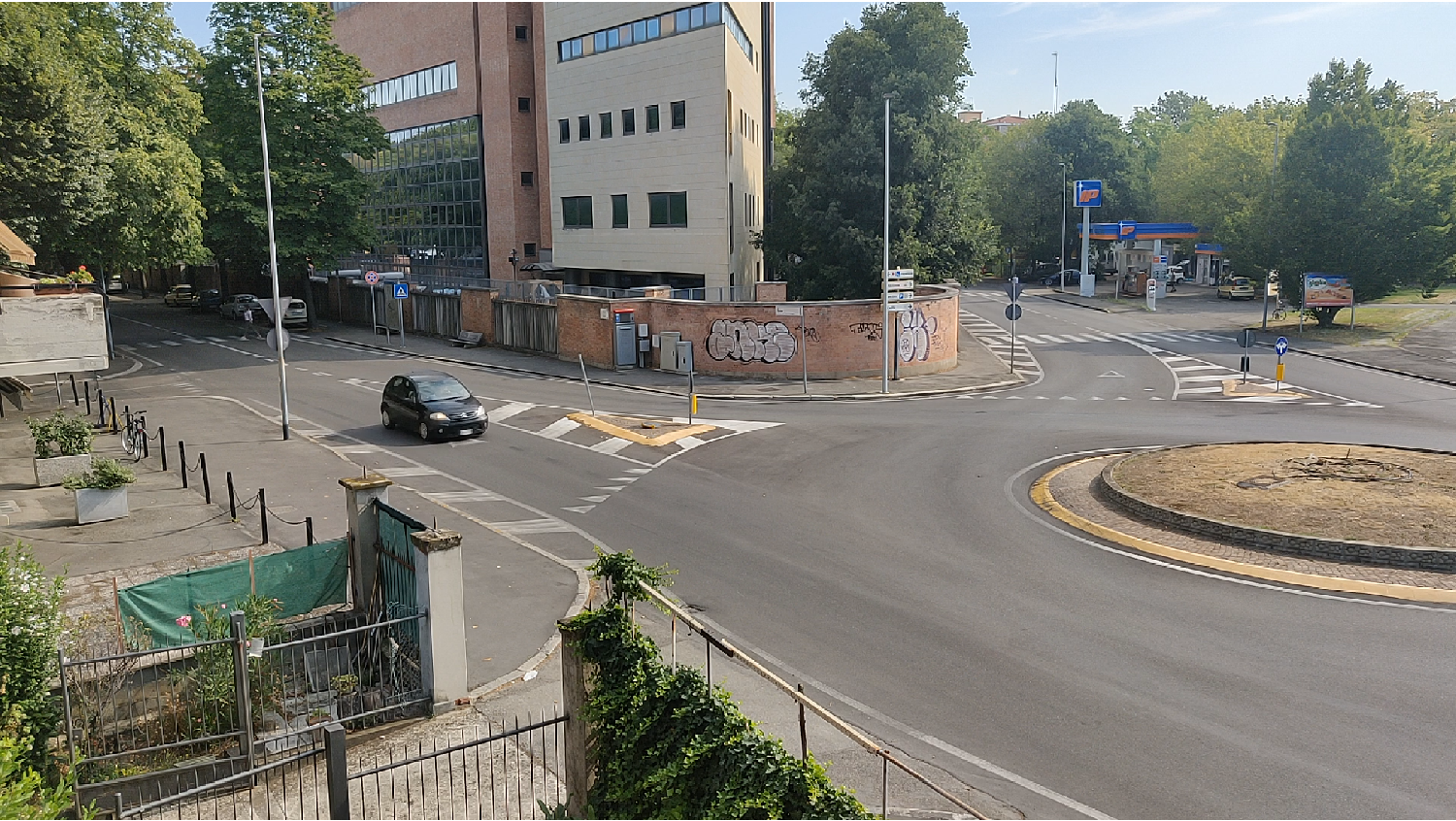} &
		\includegraphics[width = 3.9cm, height=2.4cm]{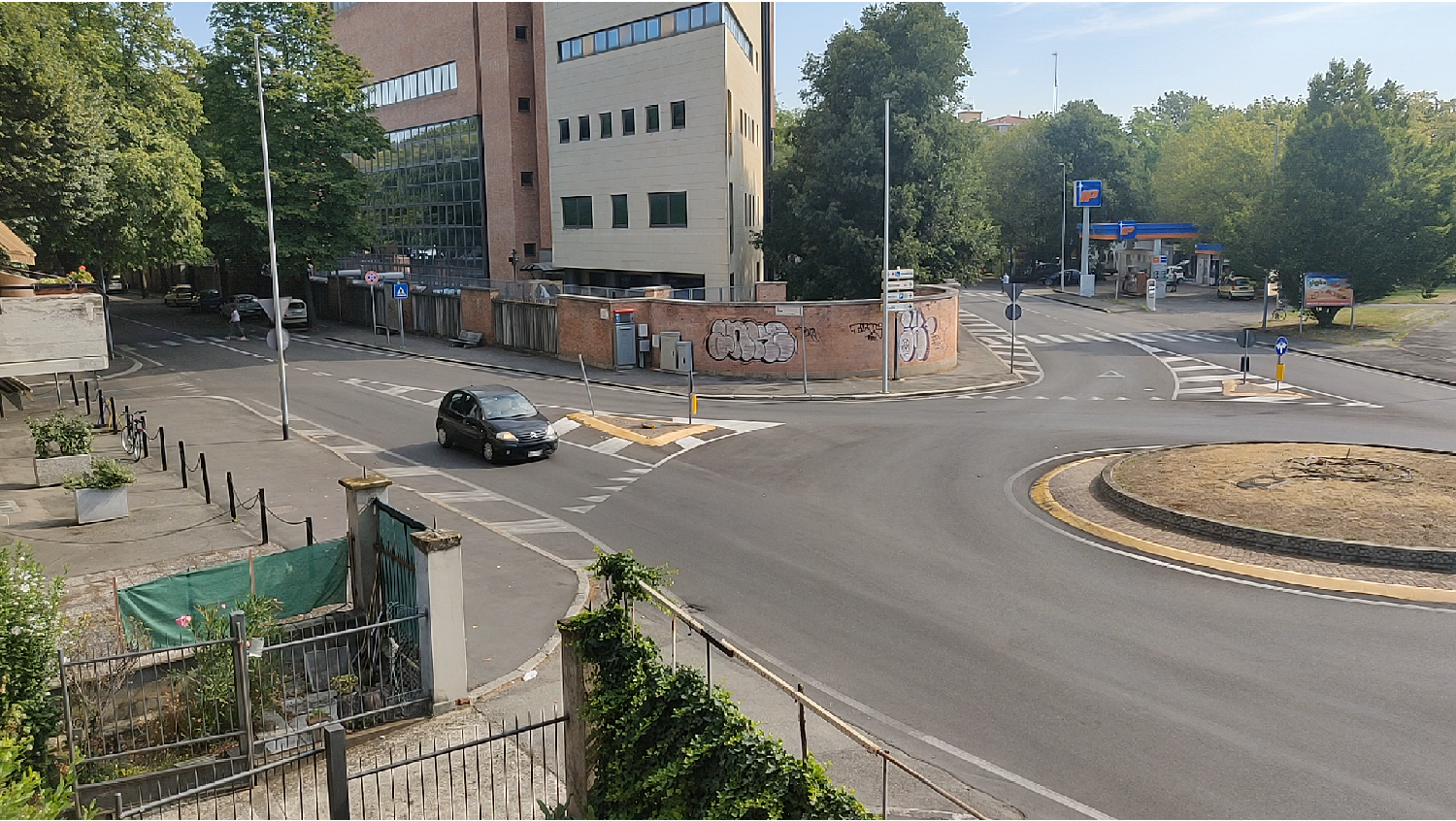} &
		\includegraphics[width = 3.9cm, height=2.4cm]{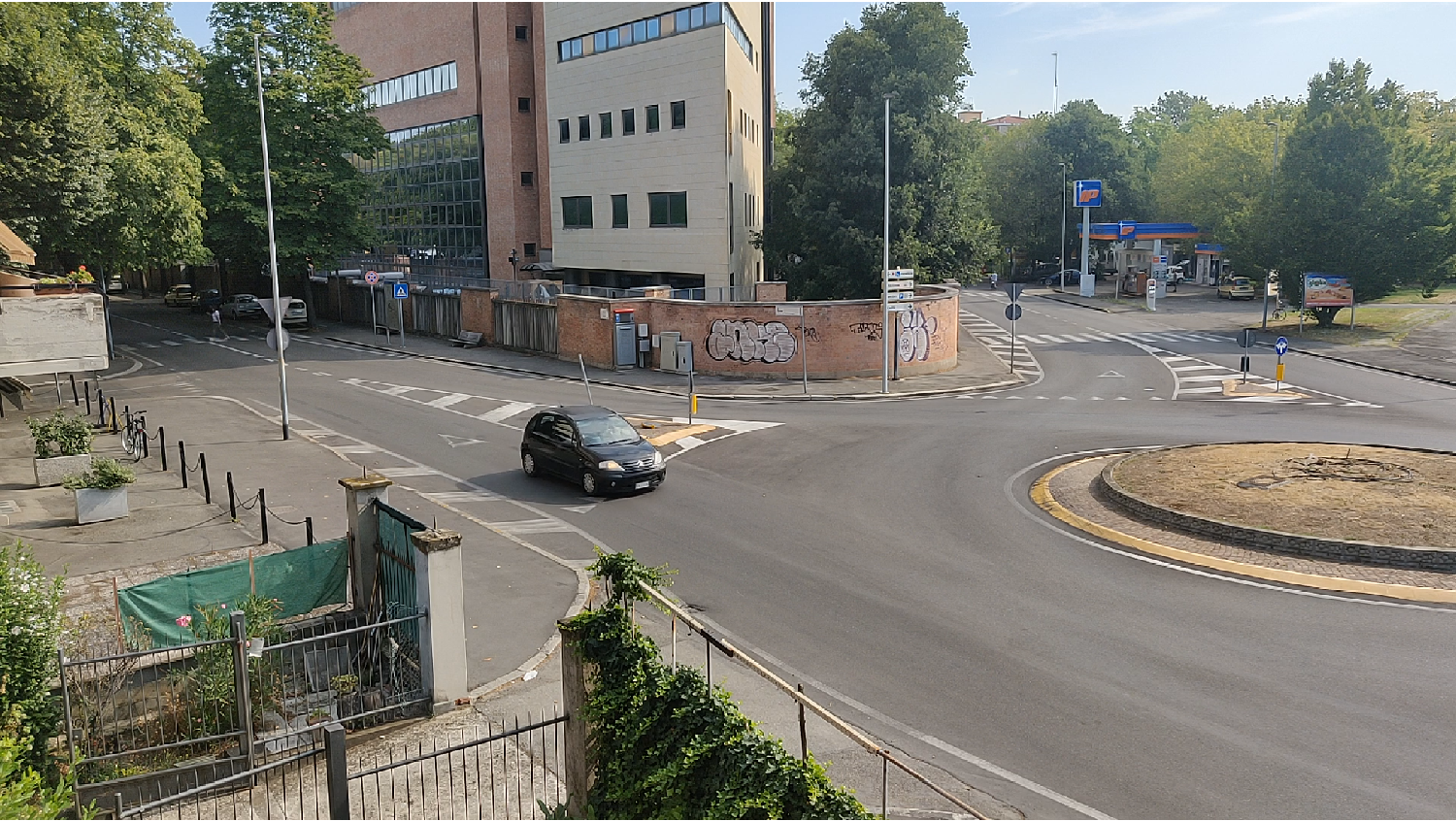} &
		(a)\\
		\includegraphics[width = 3.9cm, height=2.4cm]{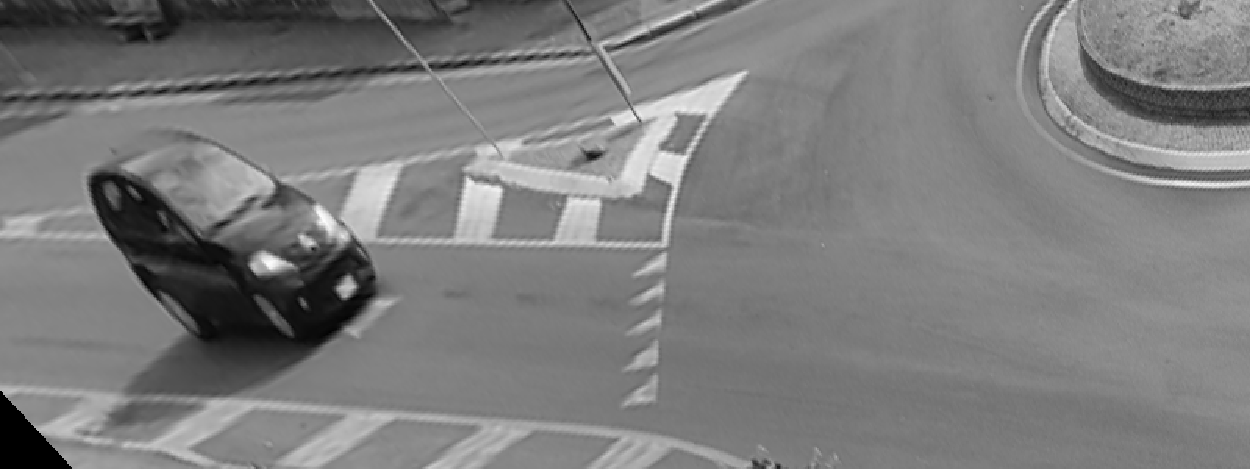} &
		\includegraphics[width = 3.9cm, height=2.4cm]{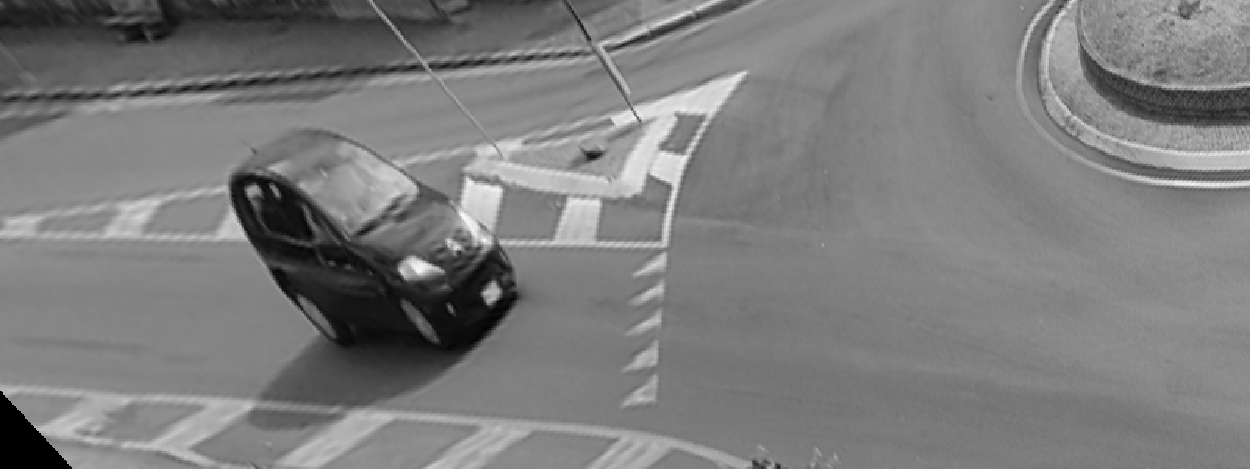} &
		\includegraphics[width = 3.9cm, height=2.4cm]{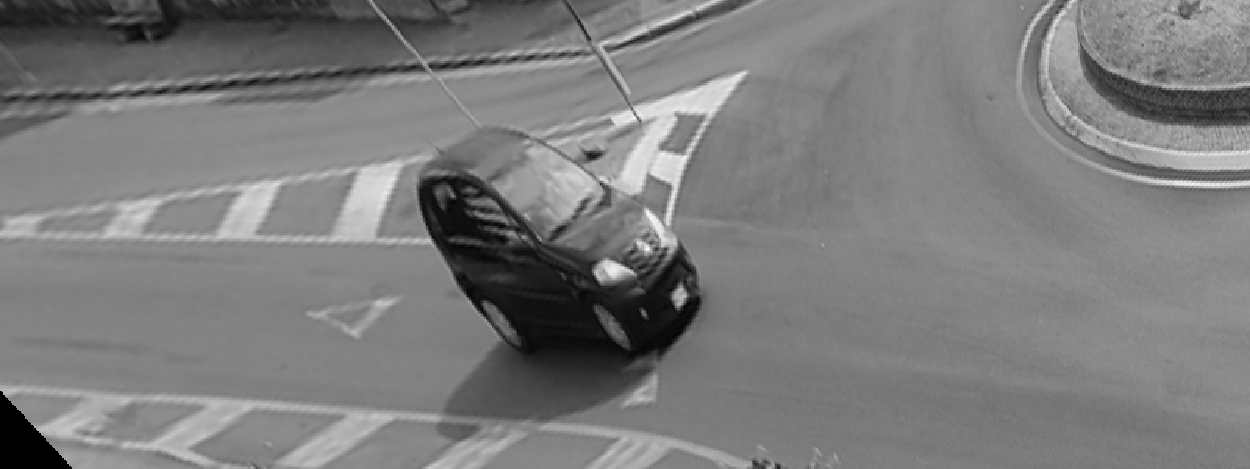} &
		(b)\\
		\includegraphics[width = 3.9cm, height=2.4cm]{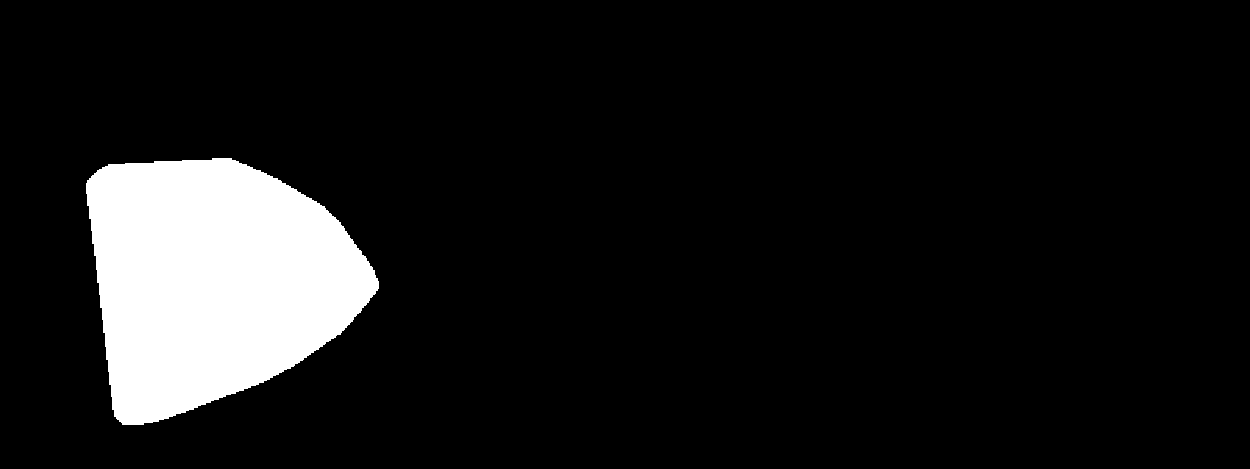} &
		\includegraphics[width = 3.9cm, height=2.4cm]{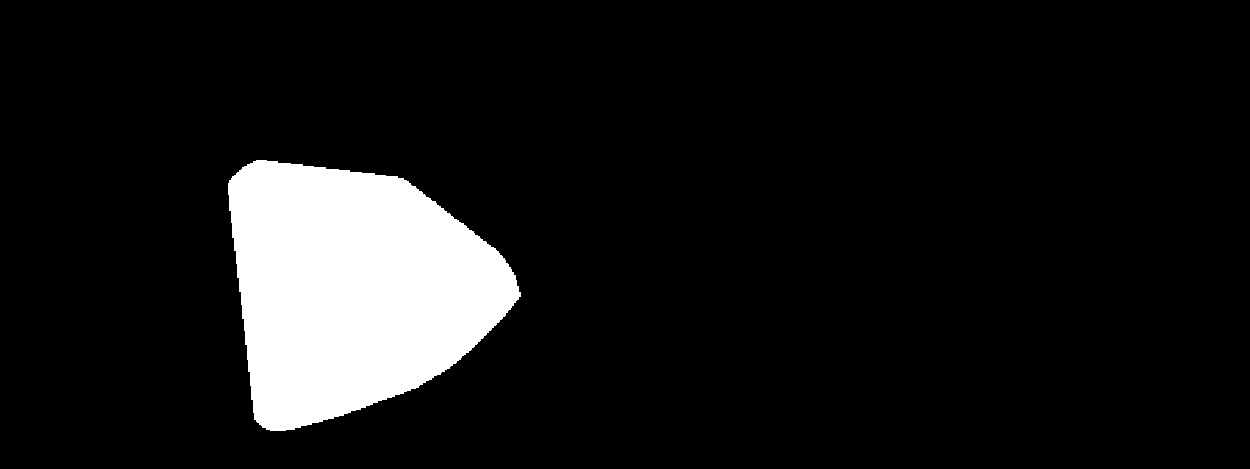} &
		\includegraphics[width = 3.9cm, height=2.4cm]{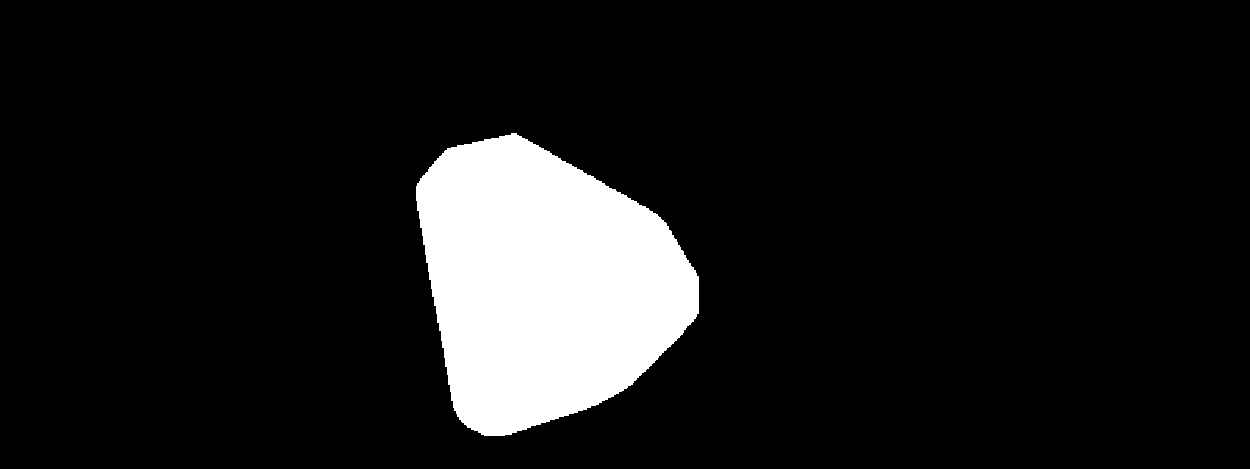} &
		(c)
	\end{tabular}
	\caption{Sample frames of: (a) original Sequence 2, (b) processed sequence after the inverse projective transformation, (c) processed sequence after background removal and convex hull extraction.}
	\label{frame2}
\end{figure*}

In Figures \ref{frame} and \ref{frame2}, a few illustrative examples of original and processed frames of two considered sequences are shown. In particular, columns correspond to different frame indices and rows (a), (b) and (c) indicate the original sequences, the processed sequences after the inverse projective transformation and those after convex hull extraction, respectively. We refer to these sequences as Sequence 1 (Figure \ref{frame}) and Sequence 2 (Figure \ref{frame2}), for brevity.

The parameters for the reference block matching method also need to be set. In particular, the block size is set to $ 75 \times 75 $ or $ 105 \times 105 $ pixel by trial and error depending on the object size. Unlike the proposed ML estimation algorithm, the block matching approach is applied to the considered processed video sequences where the background extraction and removal operations are not performed. The presence of background provides, indeed, texture information about the diversity of blocks that helps the block matching function to avoid undesired mismatches.\\
\newpage
The estimated speed values are expressed in pixel/frame and can be converted to real world measurement units, such as km/h, by camera calibration and a suitable conversion rule, e.g.,
\begin{equation}
	\hat{\mathbf{v}}_{km/h} = \hat{\mathbf{v}} \cdot \frac{l_{m}}{l_{px}} \cdot f_{s} \cdot 3.6
\end{equation}
where $l_{m}$ and $ l_{px} $ represent a reference length in the real world, expressed in meters, and its projection onto the 2D scene, expressed in pixels, respectively. The reference length $l_{m}$ could be any known element of the real world scene, e.g., the road length or width.


\subsection{Performance analysis}
\begin{figure*}[t!]
	\centering
	\begin{subfigure}{0.98\textwidth}
		\centering
		\includegraphics[width=1\textwidth]{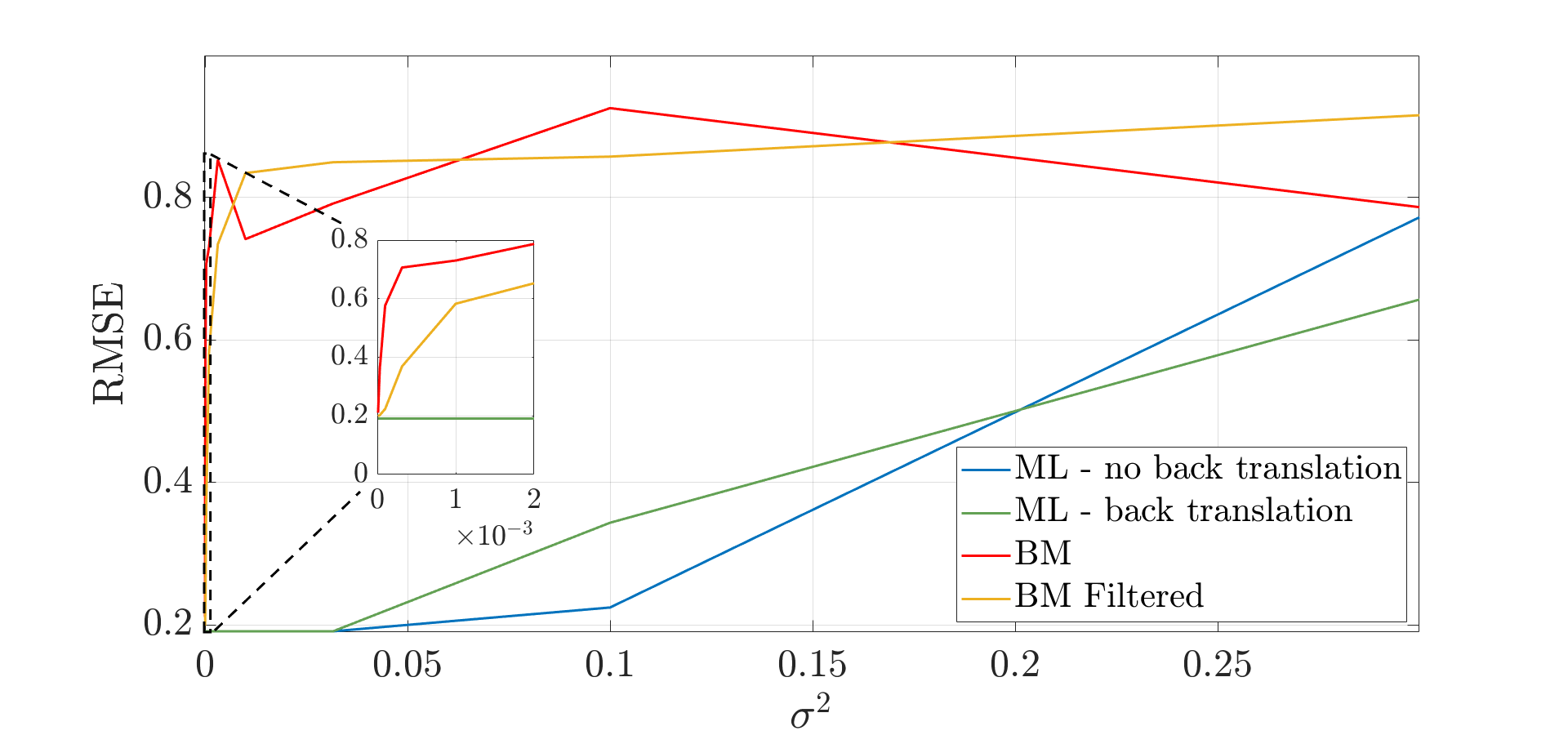}
		\caption{}
		\label{rmsef1}
	\end{subfigure}
	\\
	\begin{subfigure}{0.98\textwidth}
		\centering
		\includegraphics[width=1\textwidth]{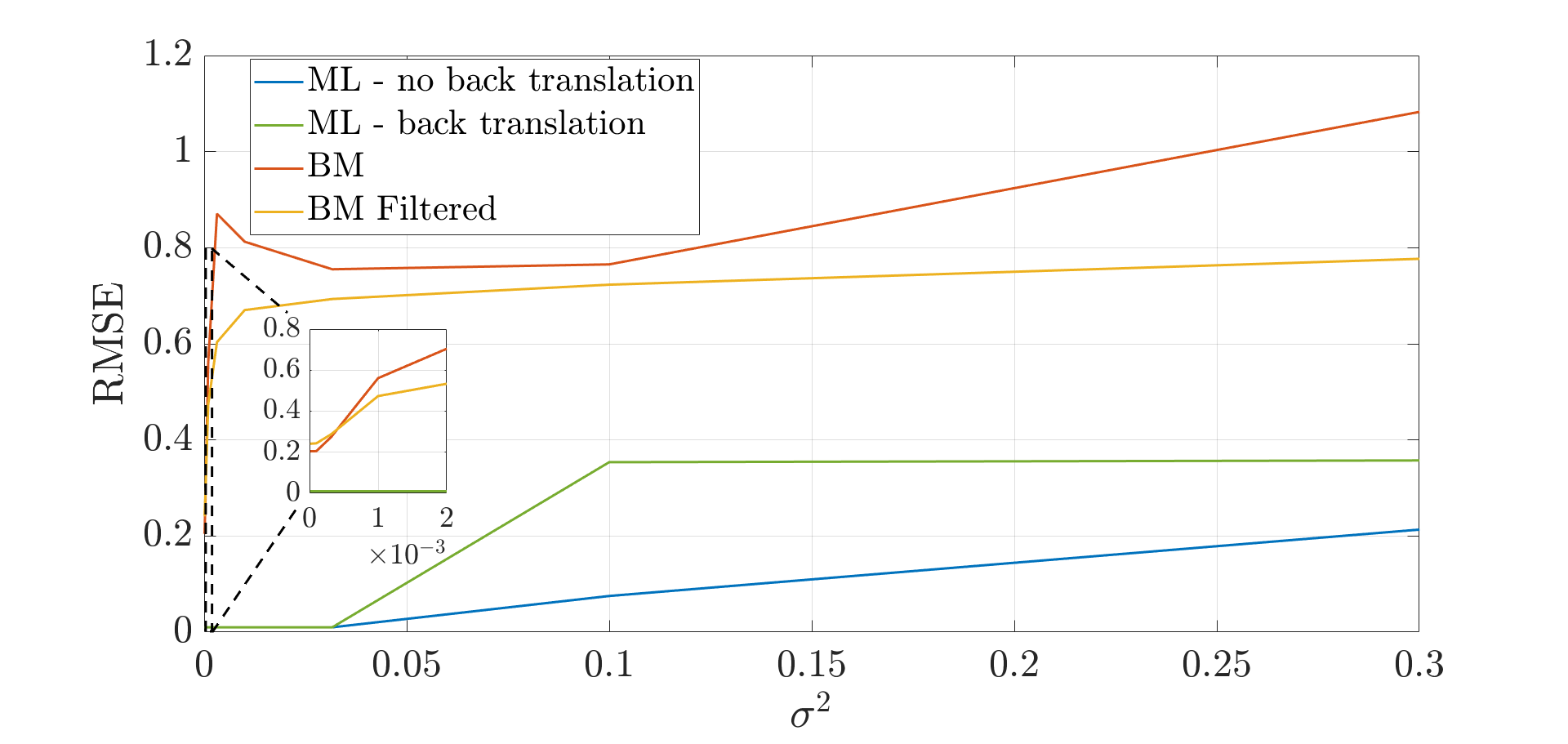}
		\caption{}
		\label{rmsev2}
	\end{subfigure}
	\caption{Performance of the assessed estimation methods in terms of RMSE vs. noise variance for: (a) sample Sequence 1 (Figure \ref{frame}) and (b) sample Sequence 2 (Figure \ref{frame2}).}
	\label{RMSE}
\end{figure*}
In Figures \ref{RMSE}(a) and \ref{RMSE}(b) the normalized RMSE in (\ref{eq_rmse}) is shown against increasing values of noise variance for the two sample sequences shown in Figures \ref{frame} and \ref{frame2}, respectively. The RMSE in (\ref{eq_rmse}) is hence computed with $J = 1$ and $R=10$. The image of the foreground moving object can be considered constant in both examples, as can be observed in the rows (c) of both Figures \ref{frame} and \ref{frame2}, where the processed sequences are shown. The ML estimation method is tested by directly maximizing (\ref{eq_J}) or (\ref{eq_J2}) and estimated speed components are searched over a quantization grid with fractional values of $ 0.5 $~pixel/frame. When the image of the object of interest can not be considered constant with respect to $n$, (\ref{eq_J}) holds and a dedicated processing operation is necessary to obtain the sequence $ S_{i}[\mathbf{k},n] $. To this purpose, the moving object of interest is translated back to its original position by shifting each frame of the sequence $ X[\mathbf{k},n] $ by the object centroid computed at the $n$-th frame after the convex hull extraction operation depicted in Figure \ref{fig1}. However, if the image of the foreground object can be considered constant, as in this case, implementing expression (\ref{eq_J}) may be unnecessary and computationally expensive. In Figure~\ref{RMSE}, results obtained by applying (\ref{eq_J}) and (\ref{eq_J2}) are referred to as ``ML~-~back translation" and ``ML~-~no back translation", respectively. Both options are here analysed for the sake of completeness and the trend of the respective RMSE curves plotted in Figure~\ref{RMSE} confirms that their performance is equivalent, especially for low values of noise variance. 

On the other hand, the block matching method is tested with and without the application of the spatial $ 7 \times 7 $ pixel average filter. These variations of the algorithm are indicated in Figure~\ref{RMSE} as ``BM~Filtered" and ``BM", respectively. According to the results in Figure \ref{RMSE}, the filter has a positive effect especially for high values of noise variance. It can also be observed that the RMSE obtained with the block-matching method does not reach zero even in the absence of noise, conversely to the ML curves in Figure \ref{RMSE}(b). In this specific case, the RMSE curves for the ML-based approaches are always far better than the ones obtained by the block-matching method. For the sake of visualisation, the rapidly increasing trend of the curves obtained with the block-matching method for low values of noise variance $\sigma^{2}$ can be better observed in the insets depicted in Figures \ref{RMSE}(a) and \ref{RMSE}(b). 

\begin{figure*}[t!]
	\hspace*{-0.7cm}
	\centering
	\includegraphics[width=0.98\textwidth]{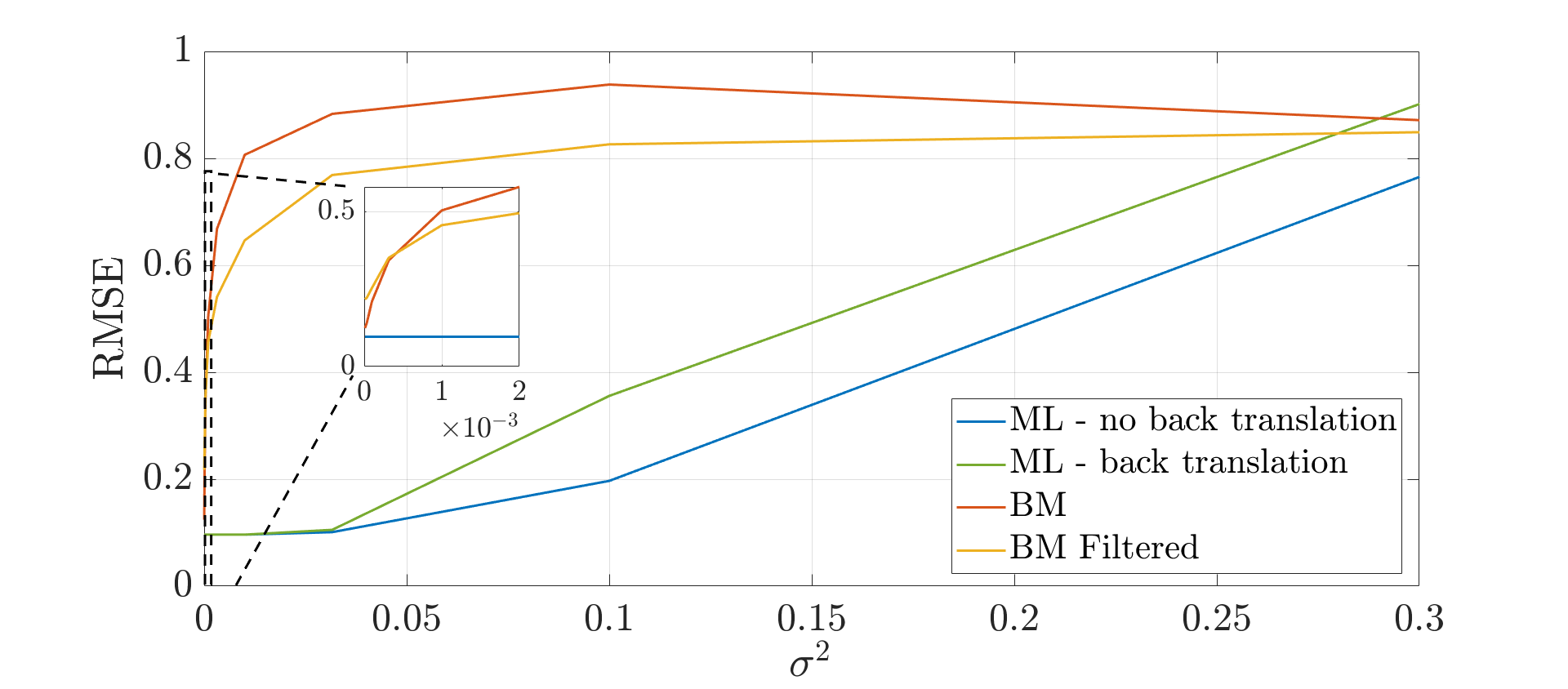}
	\caption{Performance of the assessed estimation methods in terms of RMSE vs. noise variance for a set of two video sequences where the camera angle, position and location is preserved.}
	\label{rmse_mont}
\end{figure*}

As a further example, the normalized RMSE in (\ref{eq_rmse}) is also computed for increasing noise variance $\sigma^{2}$ on a set of two video sequences where the same perspectival conditions (i.e., camera viewpoint, position and location) are preserved. In this case, $J = 2$ and $R=10$ in (\ref{eq_rmse}). The obtained results are shown in Figure \ref{rmse_mont} and confirm the performance observed in the previous Figure \ref{RMSE}.

\begin{figure*}[t!]
	\hspace*{-0.7cm}
	\centering
	\includegraphics[width=0.98\textwidth]{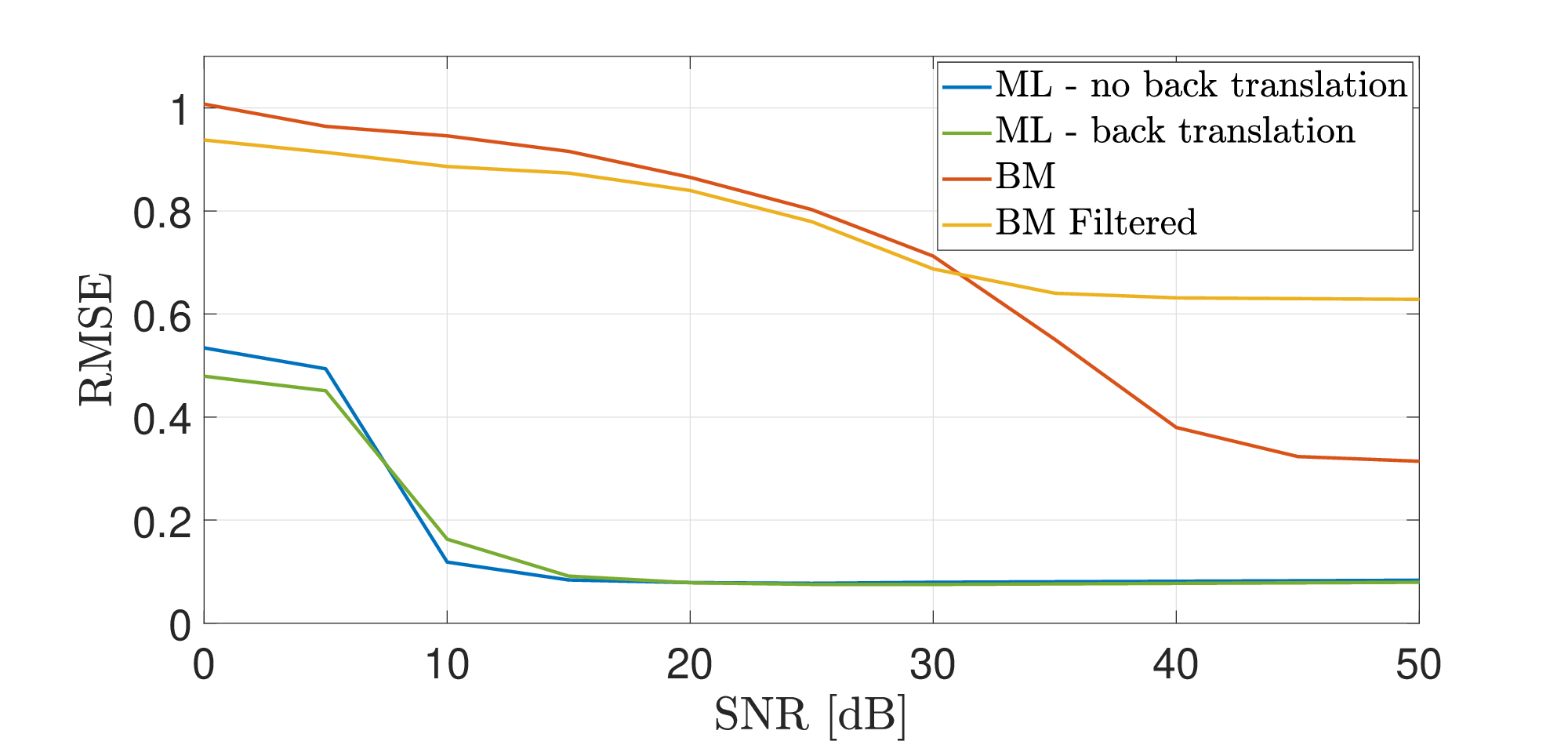}
	\caption{Performance of the assessed estimation methods in terms of RMSE vs. SNR.}
	\label{rmse}
\end{figure*}

In Figure \ref{rmse}, the normalized RMSE in (\ref{eq_rmse}) is finally shown against increasing values of the SNR, defined as $1/ \sigma^{2}$, for all considered scenarios, i.e., $J=12$ and $R=10$.
Both ML curves tend to stabilize around 0.07 because of the error introduced by the used quantization level. This value is in agreement with an estimate of normalized RMSE obtained by assuming uniformly distributed quantization error over the range $[-0.5, 0.5)$ in each dimension which, for the given video sequences, is about $0.04$.

Observing the curves obtained for the block-matching method, the mentioned effect of the spatial average filter is confirmed: it is slightly positive for low values of SNR, but excessively smoothing at high ones. The performance of the block-matching approach in both cases is far below the proposed ML estimation method. The presence of noise, even when comparatively low, prevents indeed the block-matching algorithm from correctly detecting and matching blocks. The block size and repetitive patterns, such as road lines, which are present in some of the analysed scenarios, are critical aspects which impair significantly the overall performance of the block-matching algorithm. 

The effectiveness of the proposed ML estimation method with respect to the block-matching approach is thus demonstrated in the considered heterogeneous set of realistic videos accounting for different perspectival views. Thanks to sound pre-processing operations, the presented method is robust against noise, achieving low values of RMSE also for low values of SNR and high values of noise variance $\sigma^{2}$.


\section{Conclusion} \label{conclusion}
In this work a novel method to estimate the speed of foreground objects in video signals is proposed. A model to describe the motion of objects undergoing dynamic changes, such as perspectival transformations, is derived, proper pre-processing operations are defined and the ML principle is applied to obtain an estimator of the speed of the framed objects. In particular, this paper aims to contribute to fill a gap, currently present in the literature on video content extraction, by applying sound estimation methods, such as the ML principle, to the context of motion analysis.

The proposed method is composed of robust video pre-processing stages followed by the speed estimation algorithm. Its performance is investigated by comparison with the well-known block matching approach, that is subject to some major limitations. Numerical validations are performed on the assessed algorithms, which are tested on real video sequences, also in the presence of noise. The tested video sequences differ in camera viewpoint, position and location in order to include in the analysis scenarios affected by different perspectival transformations. The effectiveness of the proposed method is finally analysed on a number of experimental videos demonstrating its good and robust performance.



	
  \bibliographystyle{elsarticle-num} 
  \bibliography{cas-refs}





\end{document}